\documentclass[sigconf]{acmart}
\usepackage{geometry}
\usepackage[utf8]{inputenc}
\usepackage[english]{babel}
\usepackage{booktabs} 
\usepackage{etoolbox}
\usepackage[figurename=Fig.,font={small},labelfont={bf,up}]{caption}
\usepackage[linesnumbered,ruled,vlined]{algorithm2e}
\usepackage{color, colortbl}
\usepackage{longtable}
\usepackage{xcolor}
\usepackage{soul}
\usepackage{comment}
\usepackage{url}
\usepackage{graphicx}
\usepackage{subfigure}
\usepackage{multirow}
\usepackage{textcomp}
\usepackage{lipsum}
\usepackage{tikz}
\usepackage{hyperref}
\usepackage{cleveref}
\usepackage{float}

\title{OESense: Employing Occlusion Effect for In-ear Human Sensing}

\newcommand{\SysName}{OESense }

\author{Dong Ma}
\affiliation{\institution{University of Cambridge}\country{}
}
\email{dm878@cam.ac.uk}
\author{Andrea Ferlini}
\affiliation{\institution{University of Cambridge}\country{}}
\email{af679@cam.ac.uk}

\author{Cecilia Mascolo}
\affiliation{\institution{University of Cambridge}	
	\country{}
	}
\email{cm542@cam.ac.uk}

\setcopyright{rightsretained}

\begin{document}
\fancyhead{}

\copyrightyear{2021}
\acmYear{2021}
\acmConference[MobiSys '21]{The 19th Annual International Conference on Mobile Systems, Applications, and Services}{June 24-July 2, 2021}{Virtual, WI, USA}
\acmBooktitle{The 19th Annual International Conference on Mobile Systems, Applications, and Services (MobiSys '21), June 24-July 2, 2021, Virtual, WI, USA}
\acmDOI{10.1145/3458864.3467680}
\acmISBN{978-1-4503-8443-8/21/07}

\begin{abstract}

Smart earbuds are recognized as a new wearable platform for personal-scale human motion sensing. However, due to the interference from head movement or background noise, commonly-used modalities (e.g.~accelerometer and microphone) fail to reliably detect both intense and light motions. To obviate this, we propose OESense, an acoustic-based in-ear system for general human motion sensing. The core idea behind OESense is the joint use of the occlusion effect (i.e., the enhancement of low-frequency components of bone-conducted sounds in an occluded ear canal) and inward-facing microphone, which naturally boosts the sensing signal and suppresses external interference. We prototype OESense as an earbud and evaluate its performance on three representative applications, i.e., step counting, activity recognition, and hand-to-face gesture interaction. With data collected from 31 subjects, we show that OESense achieves 99.3\% step counting recall, 98.3\% recognition recall for 5 activities, and 97.0\% recall for five tapping gestures on human face, respectively. We also demonstrate that OESense is compatible with earbuds' fundamental functionalities (e.g.~music playback and phone calls). In terms of energy, OESense consumes 746~mW during data recording and recognition and it has a response latency of 40.85~ms for gesture recognition. Our analysis indicates such overhead is acceptable and OESense is potential to be integrated into future earbuds.

\end{abstract}


\maketitle

\section{Introduction}
Earbuds, ear-worn wearables, have attracted growing attention from both industry and academia. 
This trend has witnessed manufacturers embedding multiple sensors in the earbuds' case to enrich their functionalities.
For example, Apple AirPods~\cite{airpodspro}, Sony WF-1000XM3~\cite{sony}, and Bose QuietControl 30~\cite{bose}, have been equipped with accelerometers for tapping interaction or multiple microphones for noise cancellation. 
On the other hand, the research community regards earbuds as a powerful personal-scale human sensing and computing platform~\cite{ferlini2019head,bui2019ebp,prakash2019stear,bedri2015stick,ando2017canalsense,martin2017ear}.
Nokia Bell Labs released eSense~\cite{kawsar2018esense}, a development platform that has further sparked the research community attention in earables sensing and computing.
By integrating sensors like PPG, barometer, and ultrasonic sensors, researchers have been devising a plethora of earable sensing applications, such as blood pressure monitoring~\cite{bui2019ebp}, facial expression recognition~\cite{ando2017canalsense}, and authentication~\cite{gao2019earecho}.

Compared to traditional wearables, earbuds possess two advantages for human sensing.
First, the human ear is an ideal position to capture various neurological, cardiovascular, and dietary signs, which promises great sensing potential for health monitoring. 
Second, earbuds are worn in the upper part of the body, which not only complements the sensing scope of smartphones/ smartwatches, but also is more robust to intensive body artifacts (e.g.~hand swing) during motion detection~\cite{prakash2019stear}.
\begin{figure*}[t]
	\centering
	\includegraphics[scale = 0.215]{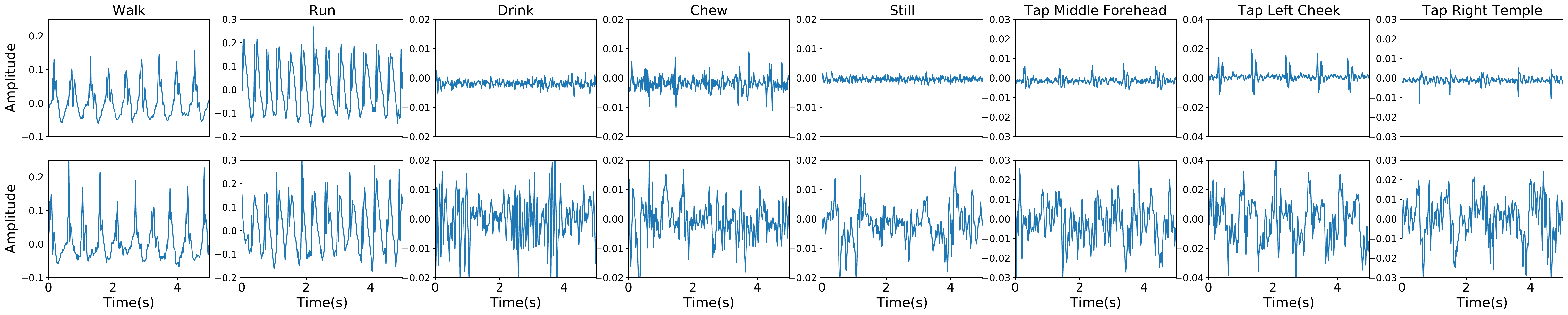}
	\caption{Comparison of signals from the accelerometer without (upper row) and with head movements (lower row). }
	\label{fig:raw_compare_acc}
\end{figure*}

\begin{figure*}[t]
	\centering
	\includegraphics[scale = 0.215]{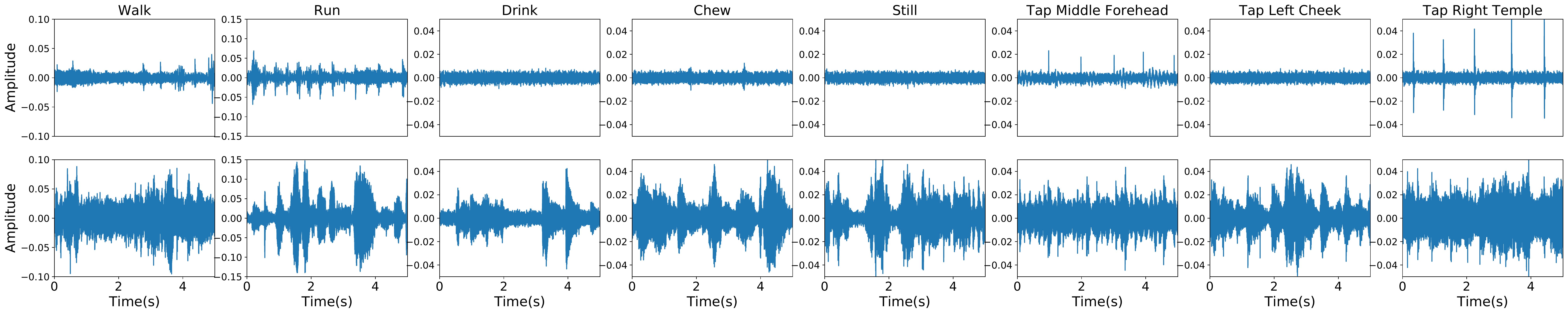}
	\caption{Comparison of signals from the external facing microphone without (upper row) and with background noise (lower row). }
	\label{fig:raw_compare_emic}
\end{figure*}

Historically, researchers used Inertial Measurement Units (IMU), accelerometers in particular, to sense motion. Some examples are human activity recognition~\cite{kwapisz2011activity}, eating habits monitoring\cite{farooq2018accelerometer}, smoking gesture recognition~\cite{parate2014risq}, and gait analysis~\cite{derawi2010accelerometer}. However, as we will show in~\Cref{sec:explore_acc}, in-ear accelerometers can only detect intense motions (e.g.~walking and running) reliably, while the signals from light motions (e.g.~chewing and tapping gestures) are compromised whenever head movements are present.
Besides accelerometers, microphones\footnote{Traditional external facing microphones.} have also been adopted to detect motion events (e.g.~gestures recognition~\cite{xu2020earbuddy}). 
However, microphone-based method suffers from low signal-to-noise ratio (SNR) due to the strong attenuation of sound in the air -- thereby limited sensing range. 
Further, they are extremely vulnerable to acoustic interference in the environment (as shown in \Cref{fig:raw_compare_emic}).

Although the sensing performance of IMUs would not be affected considerably in practice as head movements occur occasionally, some low-end earbuds or very customized hearing aids may not have IMUs on board. Thus, in the interest of form factor and cost, we explore other alternatives for human motion sensing on earables. To achieve \textit{reliable} detection of both intense and light human motions, we present OESense, a novel acoustic-based in-ear human motion sensing system. 
OEsense performs robust motion sensing based on two critical design choices.
First, to tackle environmental noise, \SysName leverages an \textit{inward-facing} microphone to record motion-induced sounds inside the ear canal.
As a result, most of the environmental noise is naturally suppressed. 
Further, acoustic signal is inherently immune to motion artifacts, like head movements. 
Second, to cope with the low SNR of traditional acoustic approaches, \SysName exploits a phenomenon known as \textit{occlusion effect} to enable the detection of both intense and light motions in human ear canal. 
Concretely, when a motion stimulus is applied to the human body, the occlusion effect boosts low-frequency bone-conducted sounds (most human motions are in a few Hertz) when the ear canal orifice is occluded. 

We prototyped \SysName with a pair of wired earbuds and a Raspberry Pi.
We selected three applications as instances of intense, mixed, and light motion detection tasks: step counting, human activity recognition, and hand-to-face gesture interactions.
We evaluated our claims with 31 subjects, demonstrating the superior sensing performance of OESense over traditional motion sensing approaches. 
Our results show OESense obtains robust performance on the three applications under various scenarios.

The contributions of this work can be enumerated as:
\begin{itemize}
\vspace{-0.05in}
\setlength\itemsep{-0.1em}
    \item We proposed the joint use of the occlusion effect and in-ear microphone for general human motion sensing, which is robust to motion/acoustic interference and capable of reliably sensing intense and light motion occurrences;
    \item We prototyped the sensing system (OESense) as an earbud, and developed sensing pipelines for three typical applications (step counting, activity recognition, and tapping gesture recognition);
    \item We proposed a software-based fit test to measure the sealing quality of earbuds, which ensures the occlusion effect is presented during motion sensing;
    \item With data collected from 31 subjects, we comprehensively evaluated the sensing performance of OESense under various conditions. Our results show that OESense achieves 99.3\% step counting recall, 98.3\% recognition recall for 5 activities, and 97.0\% recall for five tapping gestures on human face, respectively. The collected dataset was released at Kaggle~\footnote{https://www.kaggle.com/dongma878/oesense}.
    \item We assessed the system performance with power consumption and latency measurements, showing that for gesture recognition OESense consumes 746~mW power and the response latency is 40.85~ms. 
    This offers an initial indication of the feasibility of the idea which could be further optimized on more energy-efficient hardware platforms.
\end{itemize}

\section{Motivation}
\label{sec:motivation}

With this work, we aim at developing a general earable sensing system for human motion detection. The system should be able to accurately and reliably detect both intense and light (e.g.~body surface vibrations) human motions.
Specifically, we select three applications as the representatives of intense, light, and mixed motion detection tasks. 

\begin{itemize}
    \item \textbf{Step counting (intense)}: human walking involves big movements of the whole body and can be detected at different body positions (like foot, hip, waist, and head).
    \item \textbf{Human activity recognition (mixed)}: we select five activities including walking, running, being still, chewing, and drinking, which combines both intense body motions and weak surface vibrations.
    \item \textbf{Hand-to-face gesture interaction (light)}: vibrations generated by tapping different parts of the human face propagate to the ear via different paths. The received signals present distinctive patterns, enabling the recognition of different tapping gestures. This could be a potential way to interact with earables in the future.
\end{itemize}
Next, we investigate the feasibility and robustness of two commonly used sensors, i.e., accelerometer and microphone, for the three applications. 

\subsection{Preliminary Exploration: Accelerometer}
\label{sec:explore_acc}
Accelerometer has been widely adopted for motion sensing applications due to its capability of sensing both intensive and weak vibrations. It has been successfully integrated into many contemporary earbuds~\cite{airpodspro,kawsar2018esense}. However, as the human head has a high degree of freedom to move/rotate and it does not always move accordingly with the rest of body, accelerometers on the earbud are affected by head movements. To explore the severity of such interference, we embed an accelerometer (MPU6050) to an earbud and record its readings when a subject is performing eight activities defined in the three applications. 

\Cref{fig:raw_compare_acc} compares the raw acceleration signals for the eight activities, with (lower row) and without (upper row) head movements~\footnote{
The subjects randomly moved their heads in various directions as they were looking around while walking.}. We can observe that: (1) overall, the accelerometer can detect most intense (walk and run) and light (chew and taps) activities, but fails to capture extremely weak signals like drink-induced vibrations. (2) head move has minor impacts on intense activities (walk and run). The head movement only produces small variations on the signals whereas the overall patterns remain unchanged, i.e., each step is clearly observable. (3) the head movement completely obfuscates the accelerometer readings of light activities as the magnitude of head movement is larger. The above analysis clarifies the impact of head movements on the IMU data. Although head movements may not happen frequently, we found that some low-end earbuds and very customized hearing aids lack an IMU for motion sensing. Thus, we explore other modalities that are already presented in existing earbuds.

\subsection{Preliminary Exploration: Microphone}
\label{sec:explore_emic}
The microphone is the most widely available sensor in earbuds, originally used to capture human speech. Although the microphone can measure motion-induced sound thereby inferring activities, it is vulnerable to environment noise. To validate this, we record the microphone (external facing) data from an earbud when a subject performs the same activities mentioned above. 
\Cref{fig:raw_compare_emic} compares the raw microphone signals under the eight activities, with (lower row) and without (upper row) background noise (music playing). We can observe that: (1) compared to accelerometer, external-facing microphone shows less potential for motion detection (only run and two tapping gestures can be reliably detected). The reason is that external microphone measures the air-conducted sound, which suffers from strong attenuation, therefore only motions producing relatively high volume can be detected. For walking, the step sound also depends on the material of ground and shoes. (2) graphs in lower row indicate that the sensing signals are completely buried in the background music. Given that motion sounds and background music are both audible and share most of the spectrum, it would be very challenging to filter such interference with signal processing techniques. A more detailed performance comparison with accelerometer-based approach and the proposed \SysName will be presented in \Cref{sec:eval_comp}.

\section{OESense: System Design}
\subsection{Overview}
As discussed, accelerometer and external microphone based methods cannot be applied to general motion sensing due to the impact of head moves and background noise, respectively. To achieve our aim, we propose the joint use of the occlusion effect and in-ear microphone, namely OESense, to sense human motions with earbuds. When wearing the OESense earbuds, vibrations/sounds generated by motion stimuli applied to human body will propagate to the ear canal through bone conduction and be captured by in-ear microphones. With signal processing and machine learning, the signals can be used to infer the applied human motions.

\begin{figure}[ht]
	\centering
	\subfigure[]{
	\includegraphics[scale = 0.21]{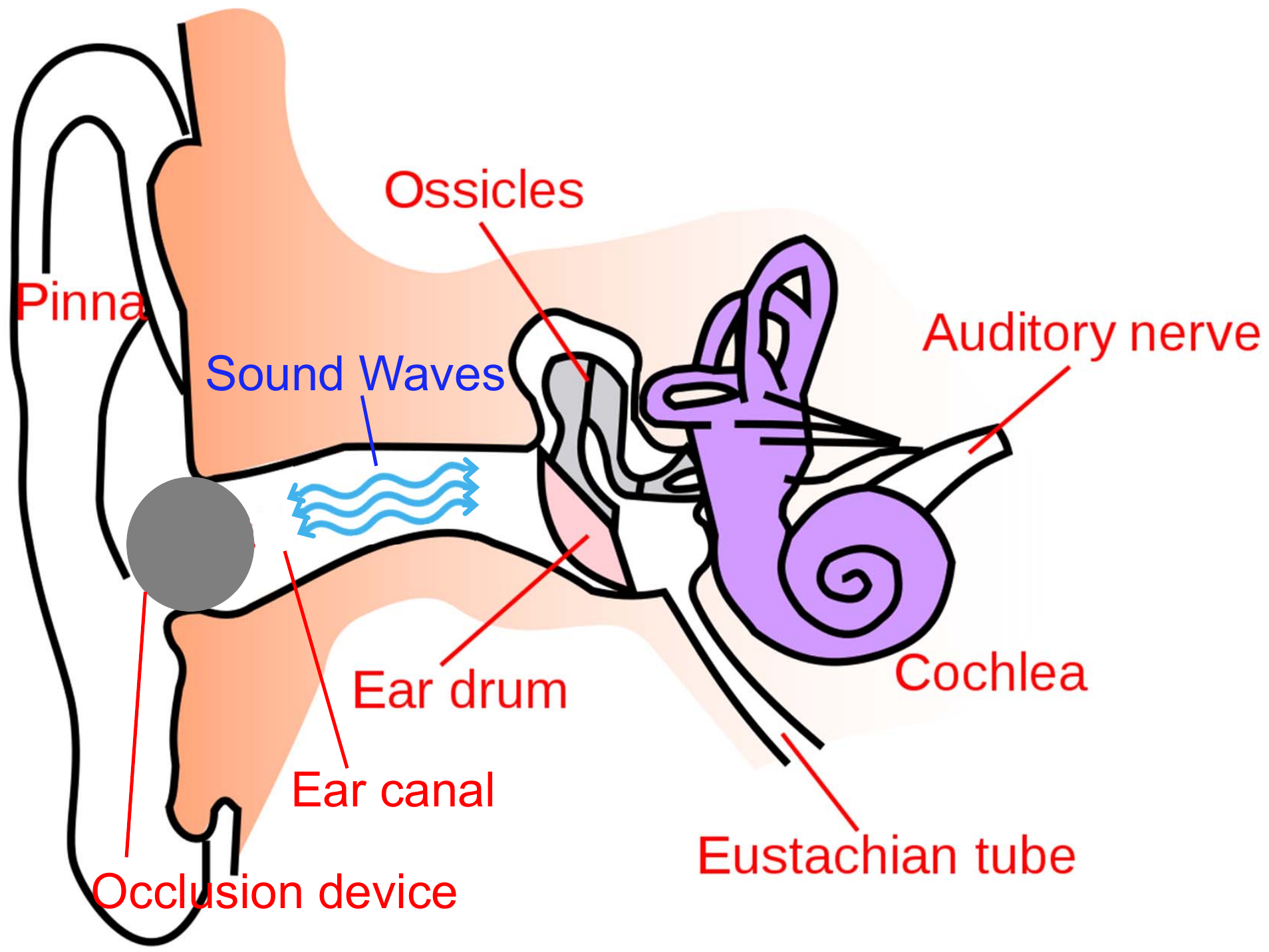}
	\label{fig:occlusion_effect}}
	\subfigure[]{
	\includegraphics[scale = 0.23]{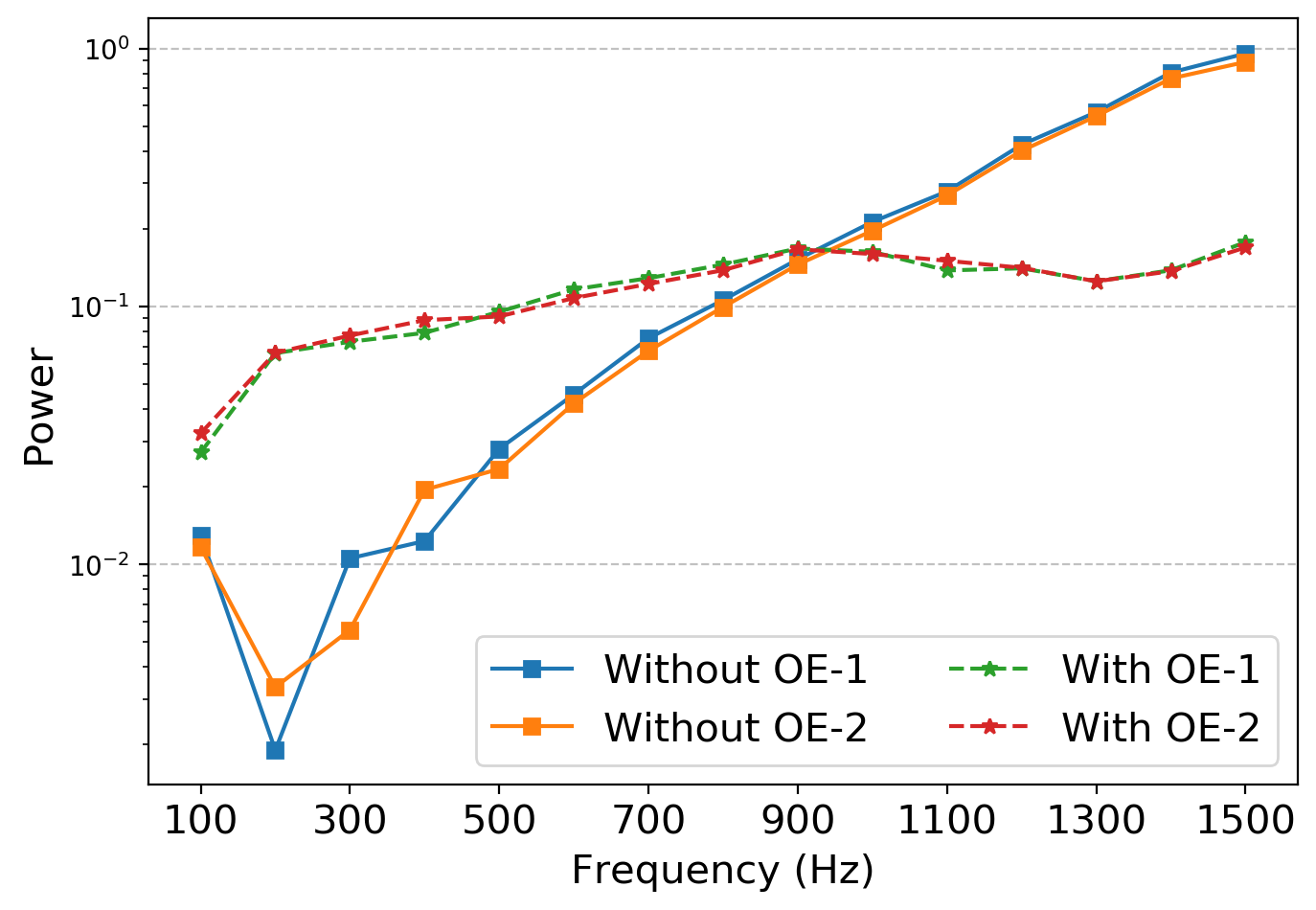}
	\label{fig:OE_demonstration}}
	\caption{(a) Illustration of human ear anatomy and the occlusion effect, (b) Impact of occlusion effect on the frequency response. 1 and 2 denote two measurements, which suggests the response is highly consistent. }
	\vspace{-0.05in}
	
\end{figure}

\begin{figure*}[t]
	\centering
	\includegraphics[scale = 0.215]{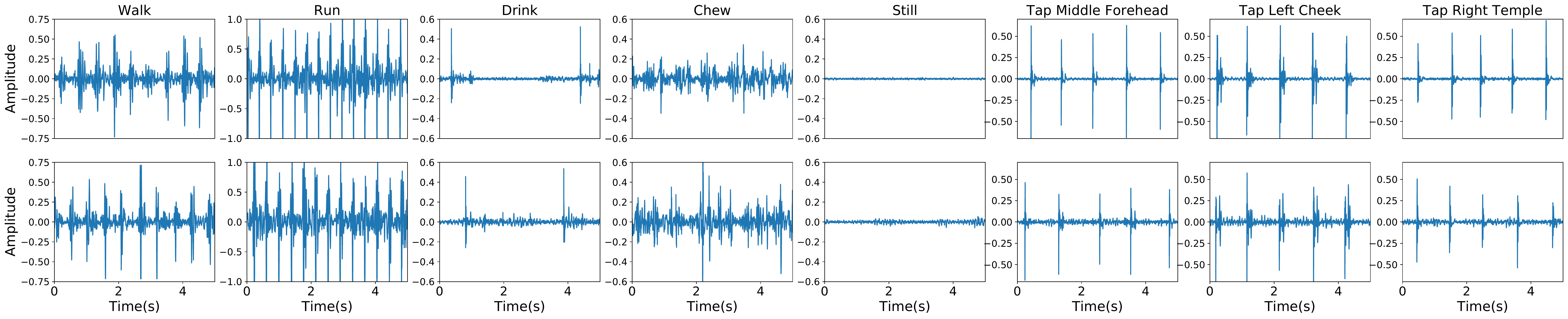}
	\vspace{-0.1in}
	\caption{Comparison of signals from the inward-facing microphone without (upper row) and with head movements (lower row). }
	\vspace{-0.05in}
	\label{fig:raw_compare_imic_hm}
\end{figure*}

\begin{figure*}[t]
	\centering
	\includegraphics[scale = 0.215]{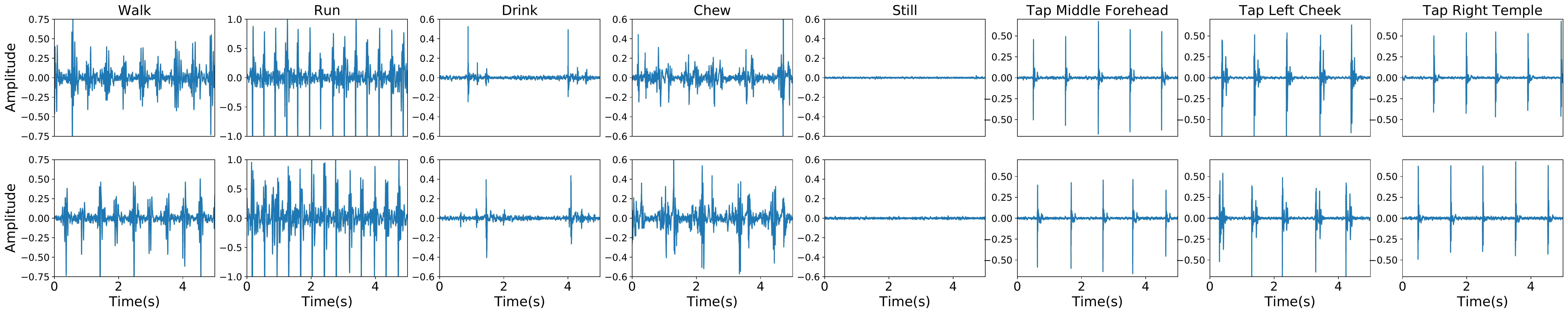}
	\vspace{-0.1in}
	\caption{Comparison of signals from the inward-facing microphone without (upper row) and with background noise (lower row). }
	\label{fig:raw_compare_imic_bn}
\end{figure*}

\subsection{Occlusion Effect}
\label{sec:occ_effect}
We now provide an explanation of the occlusion effect and how we envisage employing it for human motion sensing. When vibratory stimuli are applied on the human body, the generated sound will propagate to other parts of the body through bone conduction. Ordinarily, bone-conducted sounds induce the vibration of the ear canal wall, and the generated sounds will escape through the opening of the ear canal. However, when the ear canal is blocked, sounds are trapped and reflected back to the eardrum~\cite{stenfelt2011acoustic}, as shown in Figure~\ref{fig:occlusion_effect}. Such occlusion also increases the acoustic impedance of the ear canal opening at low frequencies~\cite{schlieper2019relationship}. Therefore, the occlusion effect is defined as the enhancement of low-frequency components of a bone-conducted sound in an occluded ear canal~\cite{stone2014technique}. A common instance of this is that people perceive echo-like sounds of their own voice when an object (like a finger) fills the outer portion of the ear canal.

Quantitatively, the occlusion effect can be denoted as the ratio between the sound pressure in the occluded ear canal and that in the open ear~\cite{stenfelt2007model}. As measured in~\cite{carillo2020theoretical}, it can boost the sound below 1000~Hz by up to 40~dB depending on the frequency. We also measure the impact of occlusion effect on the ear canal frequency response. We use the earbud speaker to transmit a single tone between 100-1500Hz (100Hz separation) and record the reflected sound with an inward-facing microphone.  \Cref{fig:OE_demonstration} compare the frequency response with and without occlusion effect (completely blocking the ear canal opening). We can see that blocked ear canal produces stronger response at frequencies below 900~Hz, while open ear canal gains much higher response at higher frequencies. In addition, we repeat the measurements twice (remove the earbud and wear it again) and the response is highly consistent, demonstrating the robustness and reliability of the occlusion effect.

Leveraging the occlusion effect for human-related sensing promises three advantages. (1) First, due to the occlusion of ear canal orifice, the inward-facing microphone mainly captures the bone-conducted sound in the ear canal and is less susceptible to environmental noises like traffic sounds and human speech. (2) Second, given that most human-produced motions are in relatively low frequencies (a few Hertz), the amplification gain provided by the occlusion effect can improve the SNR of the sensing signal. (3) Third, although earbuds are mainly used for delivery of sounds (e.g., music or phone calls) to human ear, these sounds are usually in higher frequencies so sound delivery and human sensing (under 50Hz) can coexist without mutual interference.

\begin{figure}[t]
	\centering
	\includegraphics[width = 1.0\linewidth]{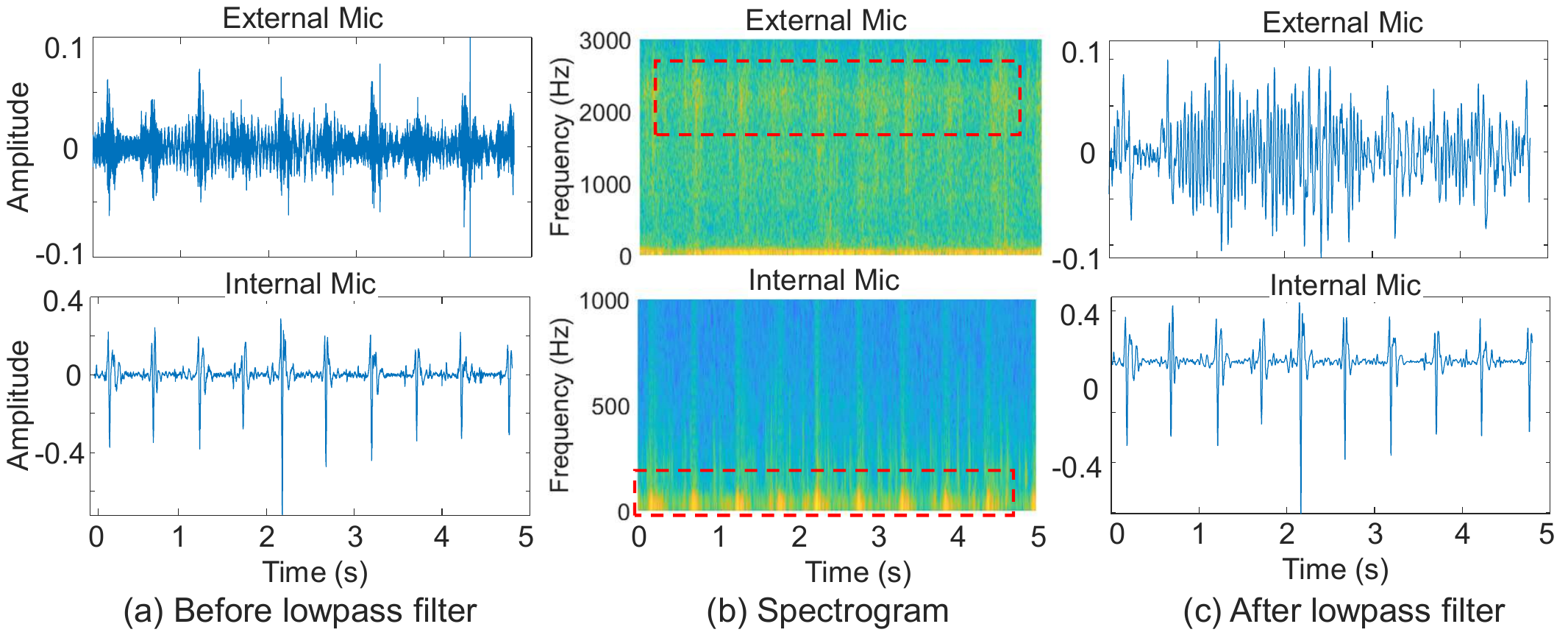}
	\caption{ Comparison of walking signals from external microphone and internal microphone, which suggests that lowpass filter does not help with the external microphone as it will remove both high frequency noise and step sounds.}
	\vspace{-0.1in}
	\label{fig:lowpass_impact}
\end{figure}

\subsection{Initial Exploration}
To demonstrate the feasibility of OESense for general motion sensing, we repeat the experiments conducted in~\Cref{sec:explore_acc} and~\Cref{sec:explore_emic} with the inward-facing microphone. \Cref{fig:raw_compare_imic_hm} illustrates the recordings for the eight activities, with and without head movements. We can see that the signals for the same activity show similar patterns under both cases, indicating head movements have negligible impact on the sensing signals.\footnote{Note that the slightly higher noise level with head movements is due to the fractions of the earbuds wires, which could be eliminated when a wireless earbud is used.} In addition, as shown in~\Cref{fig:raw_compare_imic_bn}, background noise has no impact on the sensing signals as it is naturally suppressed with the occlusion of human ear. Moreover, as the target motion signals are in frequencies below 50~Hz, any audible background noise can be easily removed with a low-pass filter. A comprehensive evaluation of the sensing performance will be presented in~\Cref{sec:eval_all}.

The fundamental difference between signals measured by external and internal microphones is the frequency band of the activity signal. As shown in~\Cref{fig:lowpass_impact} (b), the frequency of step sounds from external microphone lies between 2-3~kHz, while that for the internal microphone is below 100~Hz. Thus, although steps are observable using external microphone before lowpass filtering (\Cref{fig:lowpass_impact} (a)), the steps cannot be detected after the filtering phase (\Cref{fig:lowpass_impact} (c)). Without applying a lowpass filter, however, the external microphone would be interfered by high-frequency environment noise. Thus, lowpass filtering does not help with the external microphone as it will remove both high-frequency noise and step sounds.

\subsection{Sensing Pipelines}
After demonstrating the feasibility, we develop a lightweight sensing pipeline for each application. Specifically, we propose a robust step counting algorithm based on envelop extraction and peak detection. For activity and gesture recognition, we propose an audio-based feature set and adopt machine learning based classification.

\subsubsection{Step Counting}
\begin{figure}[t]
	\centering
	\includegraphics[scale = 0.45]{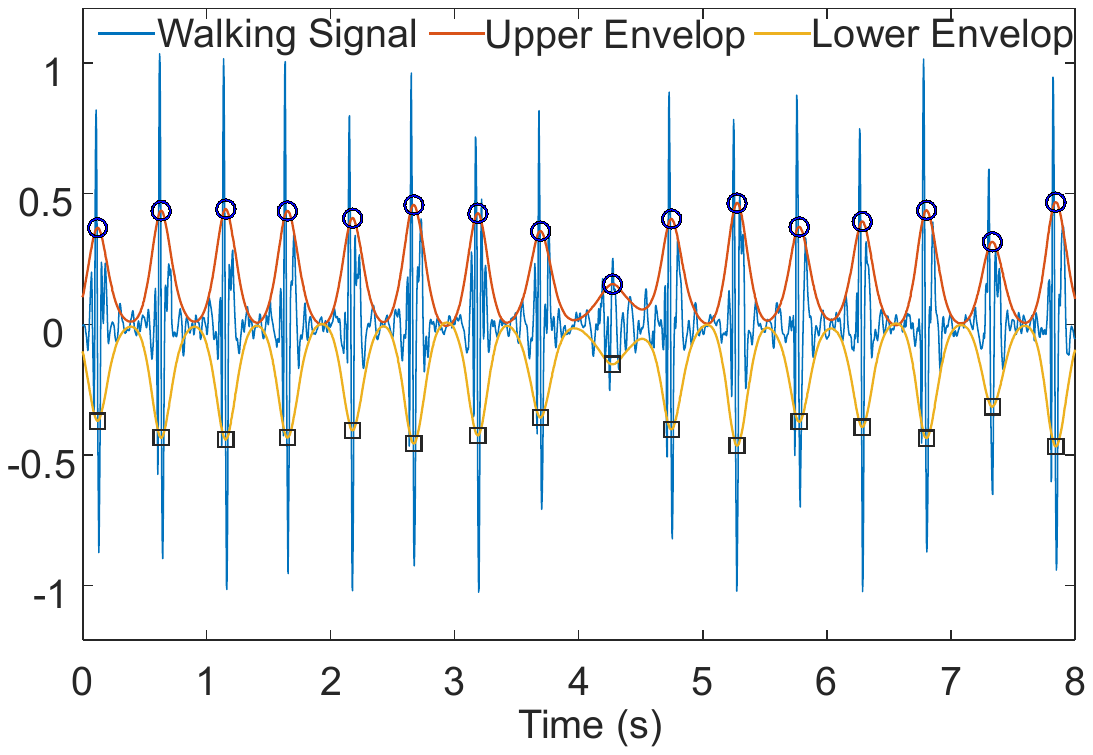}
	\vspace{-0.1in}
	\caption{A walking segment showing the performance of proposed step counting algorithm.}
	\vspace{-0.1in}
	\label{fig:step_counting_algorithm}
\end{figure}

Human steps create a periodic sinusoidal pattern on the IMU signal, so traditional step counting algorithms are to calculate the frequency of the sinusoidal signal or match with a sinusoidal template~\cite{prakash2019stear}. However, as shown in Figure~\ref{fig:step_counting_algorithm}, the step signals recorded with microphone exhibit a completely different pattern. Specifically, each step is composed of a small chunk of spikes (corresponds to foot strike stage) and a relatively silent period (corresponds to foot swing stage). Based on the observation, we proposed a step counting algorithm for audio signals. 

Given that the vibrations generated by foot strikes are at low frequency, we first apply a low-pass filter with a cut-off frequency of 50~Hz on the collected audio signal to eliminate environmental noise and human speech. Then, we feed the filtered signal $f(t)$ to the proposed algorithm. As described in Algorithm 1, the algorithm first applies the Hilbert transform on $f(t)$, which outputs the upper envelop ($up\_evlp$) and lower envelop ($low\_evlp$) of $f(t)$, as shown in Figure~\ref{fig:step_counting_algorithm}. Then, a low-pass filtered (<5~Hz) is performed on the two envelopes separately to smooth them. Afterward, we apply peak detection on the smoothed envelopes, which outputs the time index ($peak.x$) and amplitude ($peak.y$) for each peak. To avoid over-counting (i.e., false positives), we further propose two strategies to filter the detected peaks: (1) the minimum peak interval ($\theta\_intvl$) between adjacent peaks is set to 0.3~s as normal human walking frequency is lower than 3.3~Hz, (2) the minimum peak amplitude ($\theta\_amplitude$) is set to 0.3 times of average amplitude of all detected peaks. Any peak failed to satisfy either one of the conditions will be omitted. Lastly, to combat the sporadic noise that only produces an upper peak or a lower peak, we count a step only when a pair of upper peak and lower peak is aligned, i.e., the time lag (refers to maximum alignment interval $\delta$) between them is shorter than 0.2~s.

\begin{algorithm}[ht]
	\DontPrintSemicolon 
	\KwIn{Low-pass filtered signal $f(t)$, t = 1,2,...,T; minimum peak interval $\theta\_intvl$; maximum alignment interval $\delta$}
	\KwOut{Step counts $N$}
	
	$N$ $\leftarrow$ 0 \tcc{initialize N}
	
	\tcc{obtain upper and lower envelope}
	$up\_evlp(t), low\_evlp(t) \leftarrow$ HilbertTransform($f(t)$)\;
	
	\tcc{smooth envelop with low-pass filter}
	$up\_evlp'(t) \leftarrow$ LowpassFilter ($up\_evlp(t)$)\;

	\tcc{peak detection, peak=\{(x,y); x:time index, y:amplitude\}}
	$peak\_up$ $\leftarrow$ PeakDetection($up\_evlp'(t)$)\;

	\tcc{filter peak\_up and peak\_low}
	$\theta\_amplitude \leftarrow 0.3 * average(peak\_up.y)$\;
	
	\For{$i=0$;\ $i<len(peak\_up)-1$;\ $i = i+1$}{
		\If{peak\_up(i+1).x-peak\_up(i).x<$\theta\_intvl$ }{
			delete $peak\_up(i+1)$\;
		}
		\If{peak\_upper(i).y < $\theta\_amplitude$}{delete $peak\_up(i)$\;
		}
	}	
	
	\tcc{Repeat lines 3-10 for low\_evlp(t)}
	$peak\_low \leftarrow low\_evlp(t)$ 

	\tcc{Align peak\_up and peak\_low}
	\For{upper in peak\_up}{
		\For{lower in peak\_low}{
			\If{$\left| upper.x - lower.x \right|$ < $\delta$}{
				$N \leftarrow N+1$
			}
		}
	}
	
	\Return $N$
	\caption{Step counting}
	
	\label{algo:sc}
\end{algorithm}

\subsubsection{Human Activity Recognition}
\label{sec:har_pipeline}
We first apply a low-pass filter with cut-off frequency of 50~Hz on the original signal to eliminate environmental and human sounds. Then, the recorded audio stream is divided into small segments using the sliding window technique. The window size is fixed at one second with an overlapping ratio of 50\%.  Afterward, we use a widely-adopted python package called \textit{librosa}~\cite{librosa} for audio feature extraction. Specifically, inspired by~\cite{brown2020exploring}, for each instance (samples in a window), we extract 187 features that cover frequency-based, structural, statistical,
and temporal attributes. These features are Mel-frequency cepstral coefficients (MFCC) (40 features),  first-order derivative of MFCC (40), second-order derivative of MFCC (40), mel spectrogram (40),  chroma of short-time Fourier transform (STFT) (12), contrast of STFT (7), tonnetz (6), RMSE (Root Mean Square Error) (1), and onsets (1). The detailed description of these features can be found in~\cite{brown2020exploring}. 

Finally, we perform classification with five typical machine learning classifiers: Logistic Regression (LR), Support Vector Machine (SVM), K Nearest Neighbours (KNN), Decision Tree (DT), and Random Forest (RF). Due to the space limit, we report the results for LR and SVM only as they achieve the best performance. We run the experiment for data from the left earbud and right earbud individually. Then we create a \textit{fused} signal by concatenating the features extracted from the two earbuds. The reason to concatenate features instead of raw signals is to retain the authentic sequential information in the data. In specific, since the left and right earbud records concurrently, concatenating raw signals (and then performing feature extraction) introduces additional sequential information, while concatenating features avoid this issue as there is no sequential information among features. The code makes use of the Python scikit-learn package.

\subsubsection{Hand-to-face Gesture Recognition}
Similarly, we first apply a low-pass filter (<50~Hz) for denoising. Then, the envelope-based peak detection (as described in Algorithm 1) is utilized for gesture extraction. In detail, we obtain the envelope of each data trace and detect its peaks, where each peak corresponds to a tapping gesture. Then, a gesture start and end is derived by shifting 0.15 second backward and 0.25 second forward from the peak, respectively. Consequently, each gesture is composed of all samples within the 0.4 second. Then, we follow the same methodology as described in Section~\ref{sec:har_pipeline} for feature extraction and classification.

\section{Implementation}

\begin{figure}[t]
	\centering
	\includegraphics[scale = 0.35]{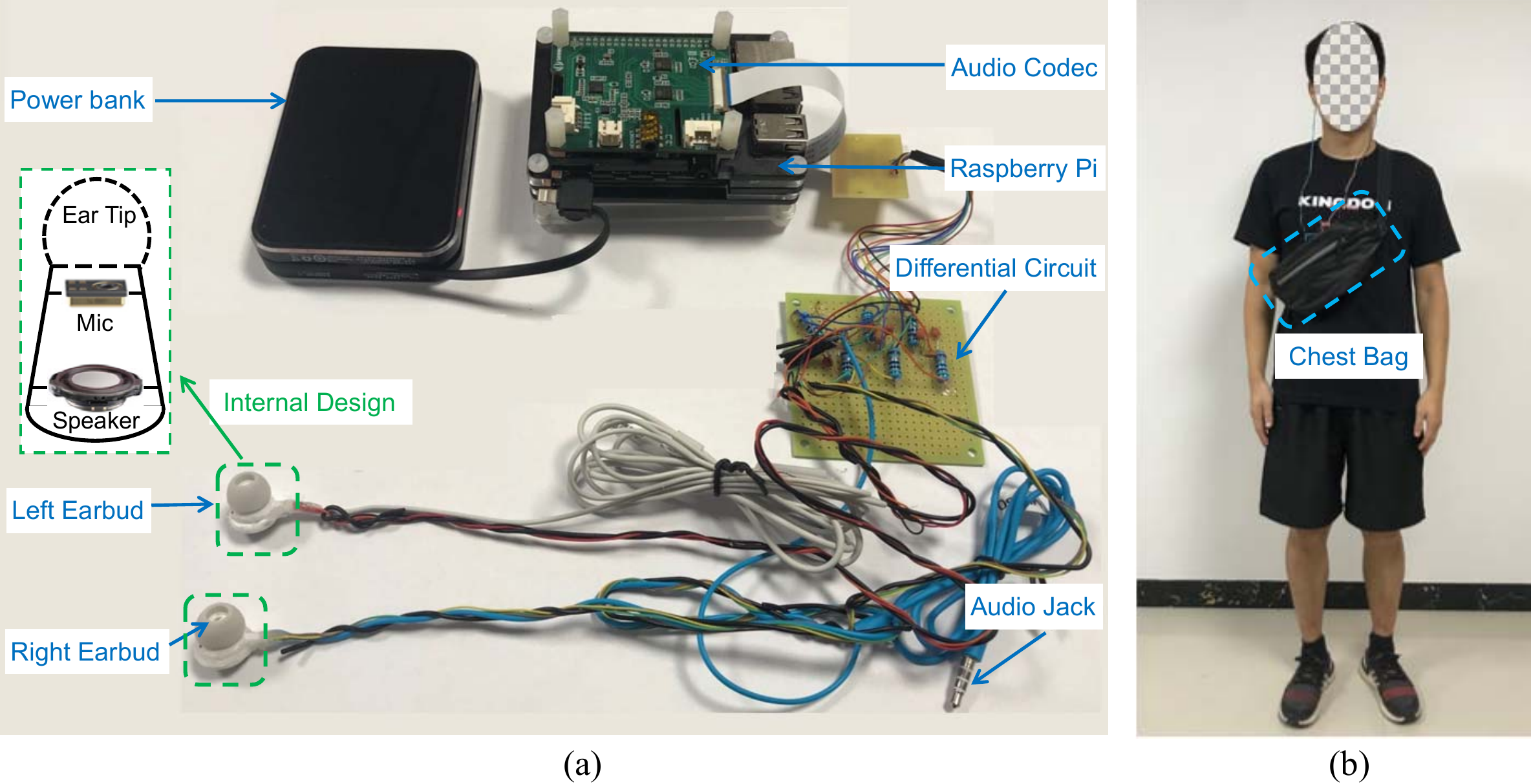}
	\vspace{-0.2in}
	\caption{(a) The developed data recording prototype, (b) illustration of a participant wearing the device.}
	\vspace{-0.2in}
	\label{fig:device_participant}
\end{figure}
\subsection{Hardware Prototyping}
Next, we present the hardware design and prototyping of OESense. Although the inward-facing microphone has been integrated into some commercial wireless earbuds~\cite{airpodspro,honormagicearbuds}, it is designed for noise cancellation and external developers cannot access the raw data. Thus, to prove our concept, we prototype \SysName by adding an inward-facing microphone to a commercial earbud. We select MINISO Marvel earphones~\cite{marvelearphones} as the base earbuds based on two criteria: (1) its internal body is large enough to accommodate a MEMS (Micro-Electro-Mechanical Systems) microphone as well as the original speaker and (2) it is equipped with silicone ear tips (removable) that can serve as the occlusion device providing good sealing quality. Then, an analog MEMS microphone (SPU1410LR5H-QB~\cite{spu1410}) is chosen to measure the in-ear sound due to its wide and flat frequency response between 20~Hz and 20~kHz. As shown in Figure~\ref{fig:device_participant} (a) (enclosed within a green dash box), we embedded the microphone at the front end of the earbud and move the original speaker to the back end. Such design optimizes the SNR of the microphone, although modifies the internal structure of the earbud. We assess whether this would affect the audio quality of music playback through user perceptions in Section~\ref{sec:dicussion}.

Besides the earbud design, Figure~\ref{fig:device_participant} (a) depicts the developed data-logger for the microphone data acquisition. A pair of earbuds are customized to measure signals in the left and right ear. To minimize noise, each microphone is connected to a differential circuit before sampled by an audio codec. We use ReSpeaker Voice Accessory HAT~\cite{repeakerhat} as the audio codec, which is controlled by a Python program running on a Raspberry Pi 4B. We sample the microphone data at 48~kHz. The speaker is connected with the original wires using a 3.5~mm audio jack. The whole prototype is powered with a power bank so that we can collect data remotely when participants are moving. To avoid affecting the subjects' walking style, all the components are enclosed in a chest bag worn by them, as shown in Figure~\ref{fig:device_participant}~(b).

\subsection{Earbud Fit Test}
The key of \SysName is the occlusion effect, which requires the ear canal to form a closed cavity by completely blocking its orifice. However, we noticed that the earbud might be loosely attached during the experiment especially in moving cases, thereby deteriorating the occlusion effect. In addition, to provide subjects with different size of silicone ear tips to ensure good sealing, we further
propose a \textit{fit test} to check the sealing quality of earbuds based on the property of occlusion effect as plotted in \Cref{fig:OE_demonstration}. 

The test works as the following: (1) before attaching the earbud to human ear, we first use the speaker to transmit two single tones of 300~Hz and 1500~Hz respectively (with a duration of 100~ms) and measure the amplitude of microphone data ($A\_300_{base}$ and $A\_1500_{base}$); (2) when the earbud is attached, repeat the single tones transmission and calculate the microphone amplitude ($A\_300_{test}$ and $A\_1500_{test}$). (3) if the amplitude ratio $A\_300_{test}/A\_300_{base}>5$ and $A\_1500_{test}/A\_500_{base}<0.2$, the sealing quality is good as the occlusion effect is observed. Otherwise, the user needs to adjust the earbud and perform the fit test again.

With the fit test, we further assess whether the occlusion effect will be reliably present when wearing the earbuds multiple times. Specifically, we asked three subjects to wear/remove the earbuds for three times (each time slightly changing the position of the earbuds), during which the system runs the fit test. The test results show that the occlusion effect exists and is consistent with normal human wearing habits, demonstrating good usability.

\section{Data Collection}
To experimentally compare the performance of \SysName with accelerometer and external microphone based approach, we first ask one subject to collect data for each application under different conditions (with and without motion and acoustic interference). The microphone and accelerometer data are sampled simultaneously at 48kHz and 100~Hz, respectively. Then, we recruited 31 participants (including 16 males, 15 females, with an age of 26.6$\pm$5.8) for large-scale data collection, during which only inward-facing microphone data is recorded~\footnote{Ethical approval for carrying out all the studies has been granted by the corresponding institution.}. 

\subsection{Step Counting}
Participants walk in a quiet meeting room (around 30 decibels) with their normal walking style and speed. The room size is 12 $\times$ 6 square meters and the participants walk in circles along the walls. To explore the robustness of step counting under various practical conditions, we consider 2 different ground materials (i.e., brick and carpet) and 5 different walking scenarios (i.e., barefoot walking, walking with slippers, walking with sneakers, walking during speaking, and walking when chewing gum). For every ground material and scenario, each participant walks continuously for 1.5 minutes, during which he/she manually counts the number of steps (serves as the ground truth)~\footnote{At normal walking speed, all the subjects walk between 156-176 steps within 1.5 minutes.}. In summary, each participant performs 10 sessions and all the subjects walk 52,047 steps in total. 

\subsection{Human Activity Recognition}
The activity experiments are conducted in the same environment as step counting. Being still, drinking, and chewing gum are performed when subjects sit on a chair. For drinking, participants hold a glass of water and keep swallowing as much as they can. For walking and running, participants move around along the room walls. In each session, participants wear the developed earbuds and perform one of the activities continuously for 1.5 minutes. Each activity is repeated two times (2 sessions), between which a 30-seconds break is set to avoid user fatigue. In total, $31 \times 5 \times (2 \times 1.5) = 465$ minutes activity data are recorded.

\begin{figure}[t]
	\centering
	\subfigure[]{
		\includegraphics[scale = 0.32]{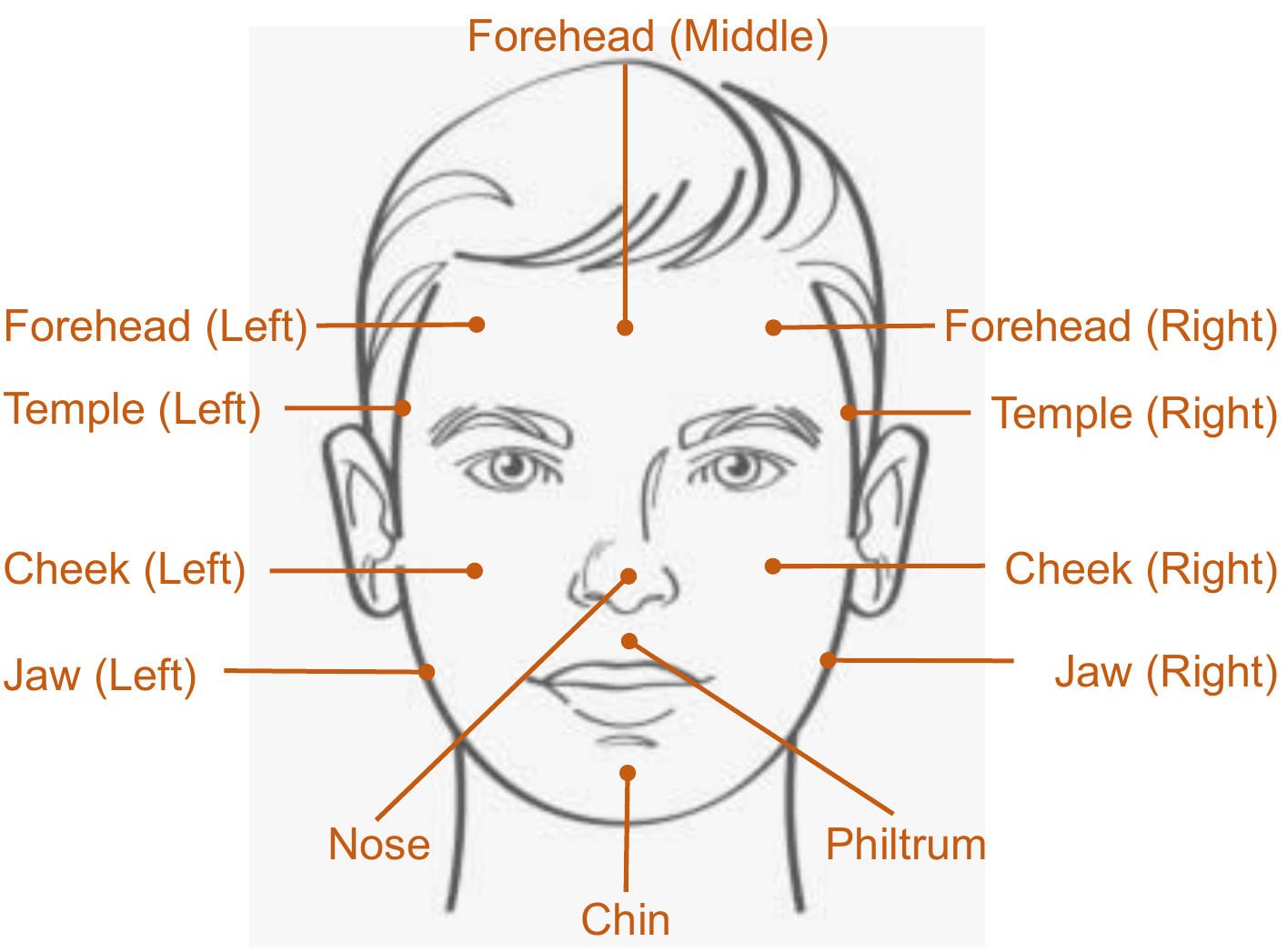}
		\label{fig:tapping_gestures}}
	\subfigure[]{
		\includegraphics[scale = 0.32]{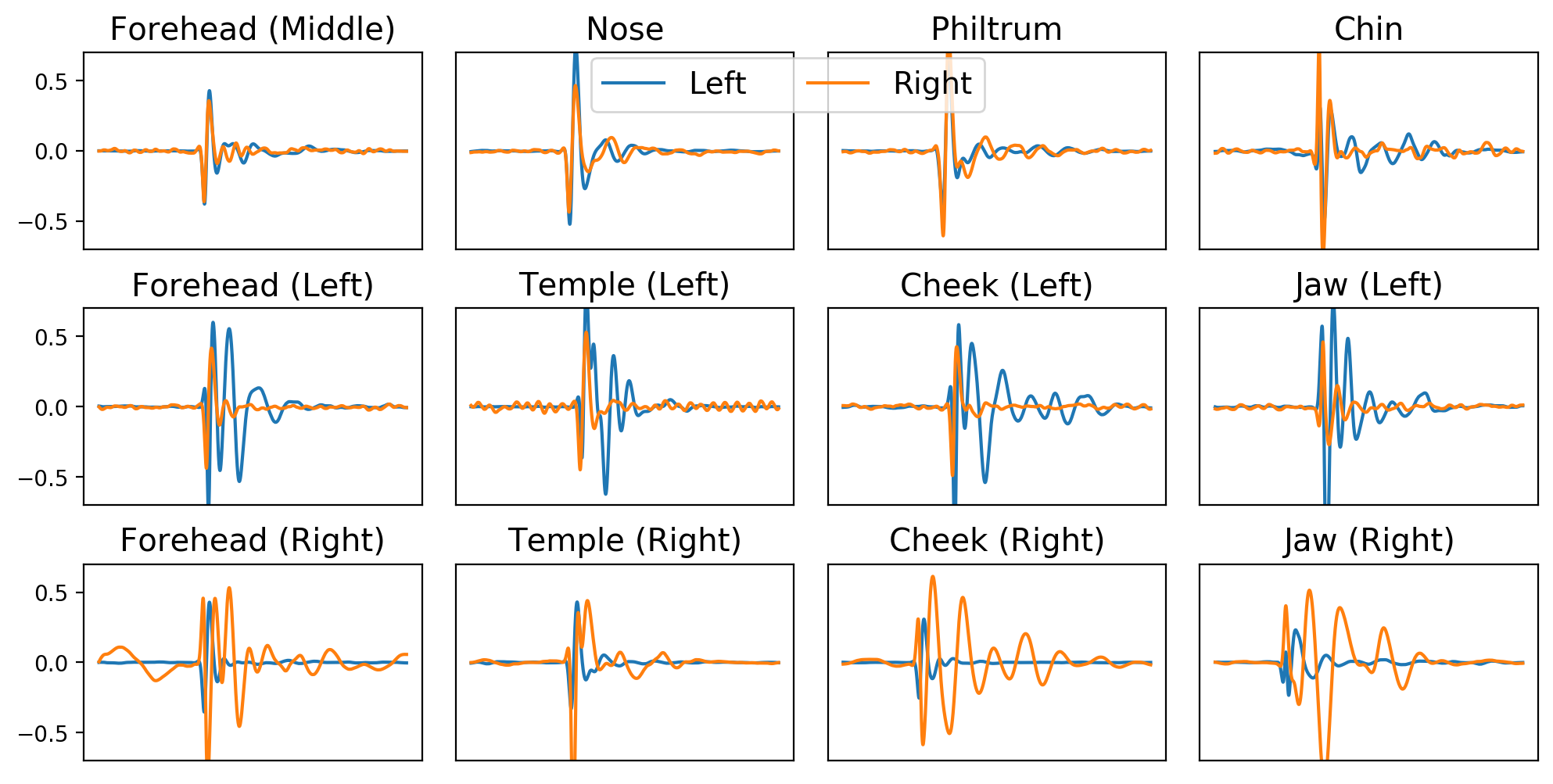}
		\label{fig:ges_waveforms}}
		\vspace{-0.2in}
	\caption{(a) Illustration of the designed tapping gestures, (b) gesture waveform from both earbuds of Subject 1.}
\end{figure}

\subsection{Hand-to-face Gesture Recognition}
As shown in Figure~\ref{fig:tapping_gestures}, we selected 12 positions (i.e., left forehead, middle forehead, right forehead, left temple, right temple, left cheek, right cheek, left jaw, right jaw, nose, philtrum, and chin) on the human face as the interaction spots. Accordingly, twelve hand-to-face gestures are created by finger tapping (one-time) on each position. Given the unique structure of bones and composition of tissues, the paths between the ear canal cavity and tapping spots are distinct, which serves as the foundation to recognize different gestures. Figure~\ref{fig:ges_waveforms} illustrates the waveforms of the 12 gestures collected from left and right earbud of Subject 1, in which we can observe that same finger taps on different spots indeed result in distinctive patterns. Further, the two earbuds can be regarded as independent sensing channels.

The participants perform each gesture 60 times with a tapping interval of 1~s. To assist the participants in maintaining the tapping interval, a cyclic one-second countdown timer is displayed on a laptop screen in front of them. All the gestures are performed with the right hand for fair comparison~\footnote{The dominant hand of all the participants is the right hand.}. The data for Subject 2 and Subject 6 are omitted as they are corrupted~\footnote{The ear tips were loosened but subjects did not report, so the occlusion effect disappeared and no gestures is detected.}. In total, we have collected $29 \times 12 \times 60 = 20,880$ gestures.

\section{Evaluation}
\label{sec:eval_all}
In this section, we first compare the performance of OESense with accelerometer and microphone based approaches (with and without interference). Then, with the large-scale dataset, we assess the performance of OESense under various conditions for each application. Lastly, we discuss the impact of music playback and the power consumption and latency of OESense.  

\subsection{Baselines Benchmarking}
\label{sec:eval_comp}
As shown in \Cref{sec:motivation}, conventional approaches, i.e.~accelerometer (Acc) and external microphone (eMic) either fail to detect light motions or suffers from motion/acoustic interference. With data collected from one subject (single ear), we run the developed sensing pipelines for each application to compare the final sensing accuracy (recall), as presented in \Cref{tab:comp_acc_mic}. For accelerometer data, we extract around 130 statistical and spectral features using the TSFEL Python library~\cite{barandas2020tsfel}. Logistic regression is used as the classifier. 

For step counting, the subject walks 300 steps in each session. We can see that accelerometer precisely counts the steps even when the interference is applied. This is because acceleration produced by head movements is negligible compared to that by walking. External microphone severely undercounts the steps in both cases as the air-conducted step sounds are very weak. OESense achieves great performance in all cases due to its immunity to motion and background noise.  

\begin{table}[t]
\centering
\caption{Performance comparison of OESense with accelerometer (Acc) and external microphone (eMic) based methods. SC-Step Counting, AR-Activity Recognition, GR-Gesture Recognition. The values for SC represent step counts (ground truth is 300), for AR and GR are recognition recall.}
\label{tab:comp_acc_mic}
\setlength\tabcolsep{5.0pt}
\begin{tabular}{lrrrr}
\toprule
&  & \textbf{ SC} & \textbf{AR} & \textbf{GR} \\ \cmidrule{1-5}
 
\multirow{2}{*}{\textbf{Acc}} &w/o head move  &300  & 72.76\%  & 59.75\% \\ \cmidrule{2-5}

&with head move  & 299  &53.01\%  & 29.55\% \\ \cmidrule{1-5}
  
\multirow{2}{*}{\textbf{eMic}} &w/o music  & 109 & 67.68\% & 46.53\%\\ \cmidrule{2-5}

&with music  & 208  & 52.47\% & 27.59\%\\ \cmidrule{1-5}

\multirow{6}{*}{\textbf{OESense}} &w/o head move  & 300  &91.26\%  &83.09\% \\ \cmidrule{2-5}

&with head move  & 299  &88.74\%  &79.62\% \\ \cmidrule{2-5}

&w/o music  & 300 & 91.26\% & 83.09\%\\ \cmidrule{2-5}

&with music  & 300  & 90.99\% &81.15\% \\ 
\bottomrule 
\vspace{-0.2in}

\end{tabular}
\end{table} 

For activity recognition and gesture recognition, we can observe that (1) without interference, Acc and eMic obtain lower recall than OEsense as they cannot detect some of the activities/gestures. (2) With interference, recognition recall of Acc and eMic decreases significantly. Overall, OESense achieves superior and comparable performance in all cases. The relatively big accuracy drop with head movements is actually caused by the fraction of earbud wires and could be resolved when a wireless earbud is developed.  

\subsection{Step counting}
\begin{figure}[t]
	\centering
	\subfigure[Brick]{
	    \centering
	    \includegraphics[scale = 0.215]{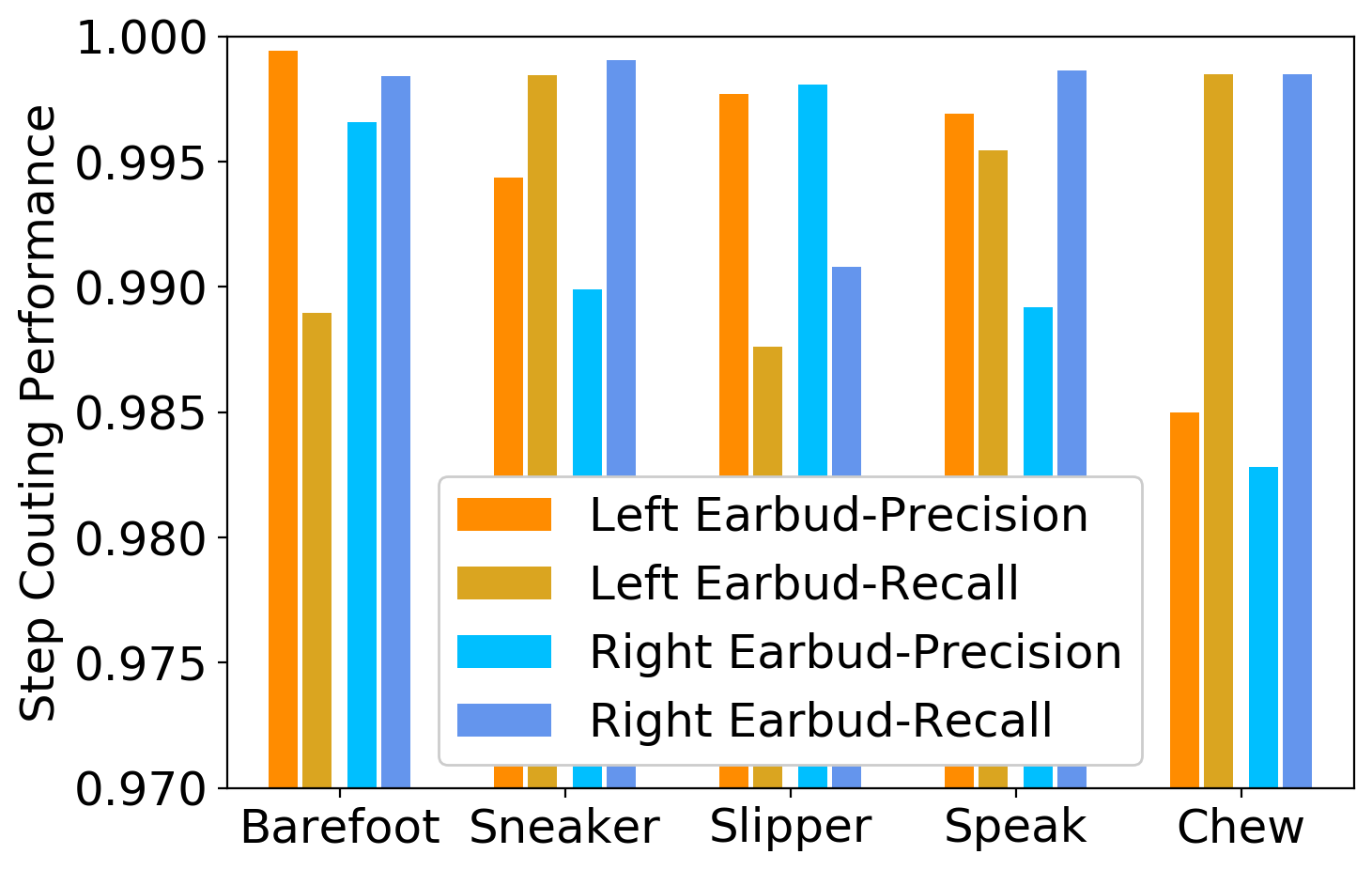}
	    \label{fig:step_brick}}
    \subfigure[Carpet]{
	    \centering
	    \includegraphics[scale = 0.215]{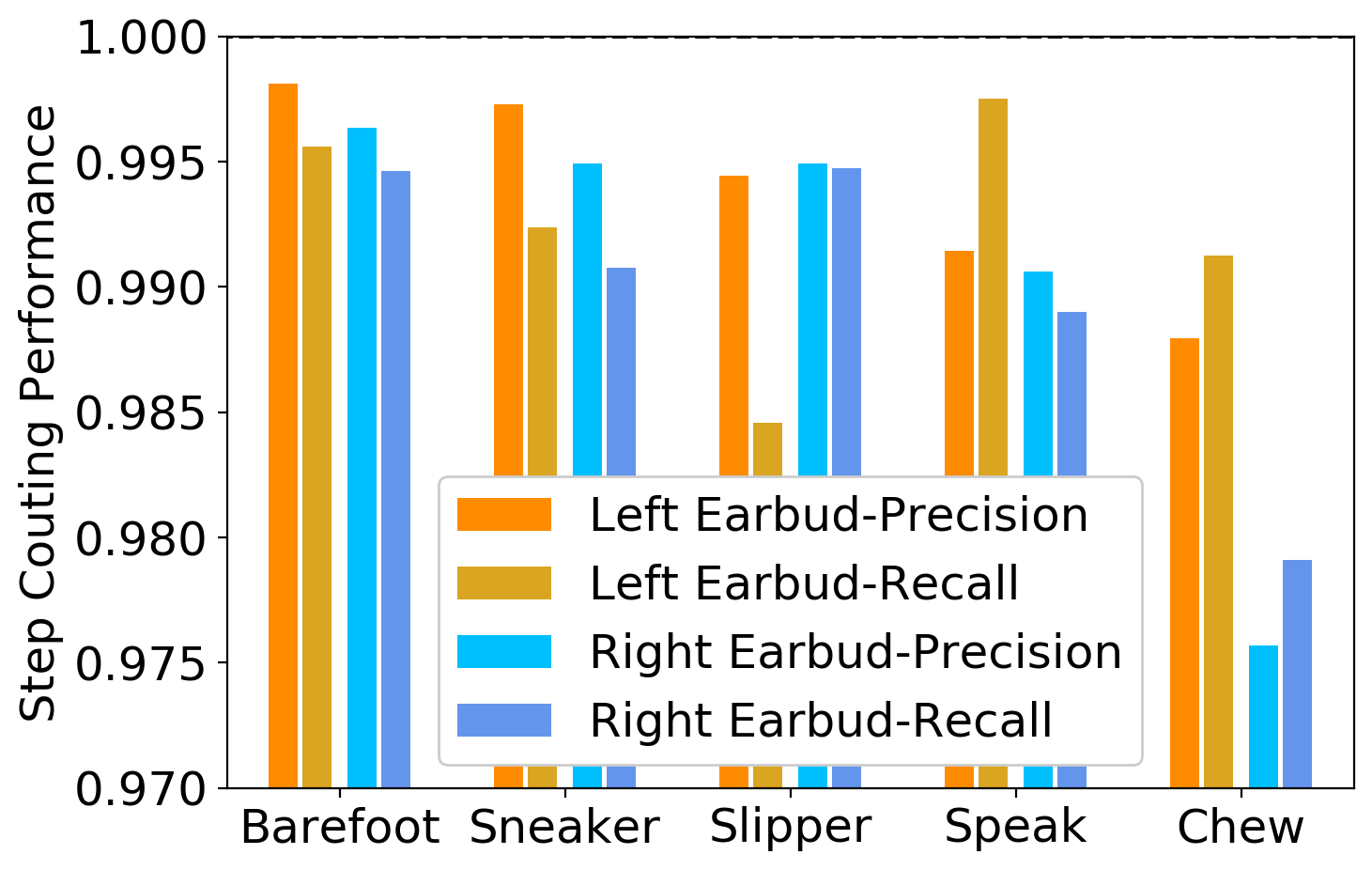}
	    \label{fig:step_carpet}}
	    \vspace{-0.2in}
	\caption{Step counting performance for walking on (a) brick and (b) carpet.}
	\vspace{-0.2in}
	 \label{fig:step_counting_result}
\end{figure}

Figure~\ref{fig:step_counting_result} depicts the step counting performance of OESense under various conditions, where precision and recall reflect over-counting and under-counting, respectively. Overall, we can observe that step counting precision and recall are higher than 97.5\% regardless of ground material and walking scenario, demonstrating the superior performance of \SysName on step counting. 

Different walking scenarios have distinct impacts on step counting performance. For example, walking with slippers will increase foot contact area during foot landing, thereby weakening the strength of the generated vibrations~\cite{yick2016effects}. So the algorithm will miss some steps (false negatives) and result in lower step counting recall. When people are chewing, vibrations will be generated by jaw movements and propagated to ear canal through bone conduction. Consequently, we can expect more spikes on the audio signal, leading to over-counting. So the precision for chewing scenario suffers the most.

For different ground materials, we can observe that brick generally obtains higher recall compared to carpet. This is because soft carpet will counteract part of vibrations and result in miss detection of steps. In addition, audio signal collected when walking on carpet has lower amplitude (i.e., lower SNR) so it is more vulnerable to other body movements. Another observation is that both the left earbud and right earbud can achieve very high accuracy, indicating the proposed step counting approach can work well even with a single earbud. Overall, \SysName achieves an average counting recall of 99.32\% and precision of 99.26\%, which dramatically outperforms the industrial standard for pedometers (for example, ±3\% counting error set by the Japanese Ministry of Economy Trade and Industry~\cite{bassett2017step}).

\subsection{Human Activity Recognition}
\begin{figure}[t]
	\subfigure[]{
		\centering
		\includegraphics[scale = 0.245]{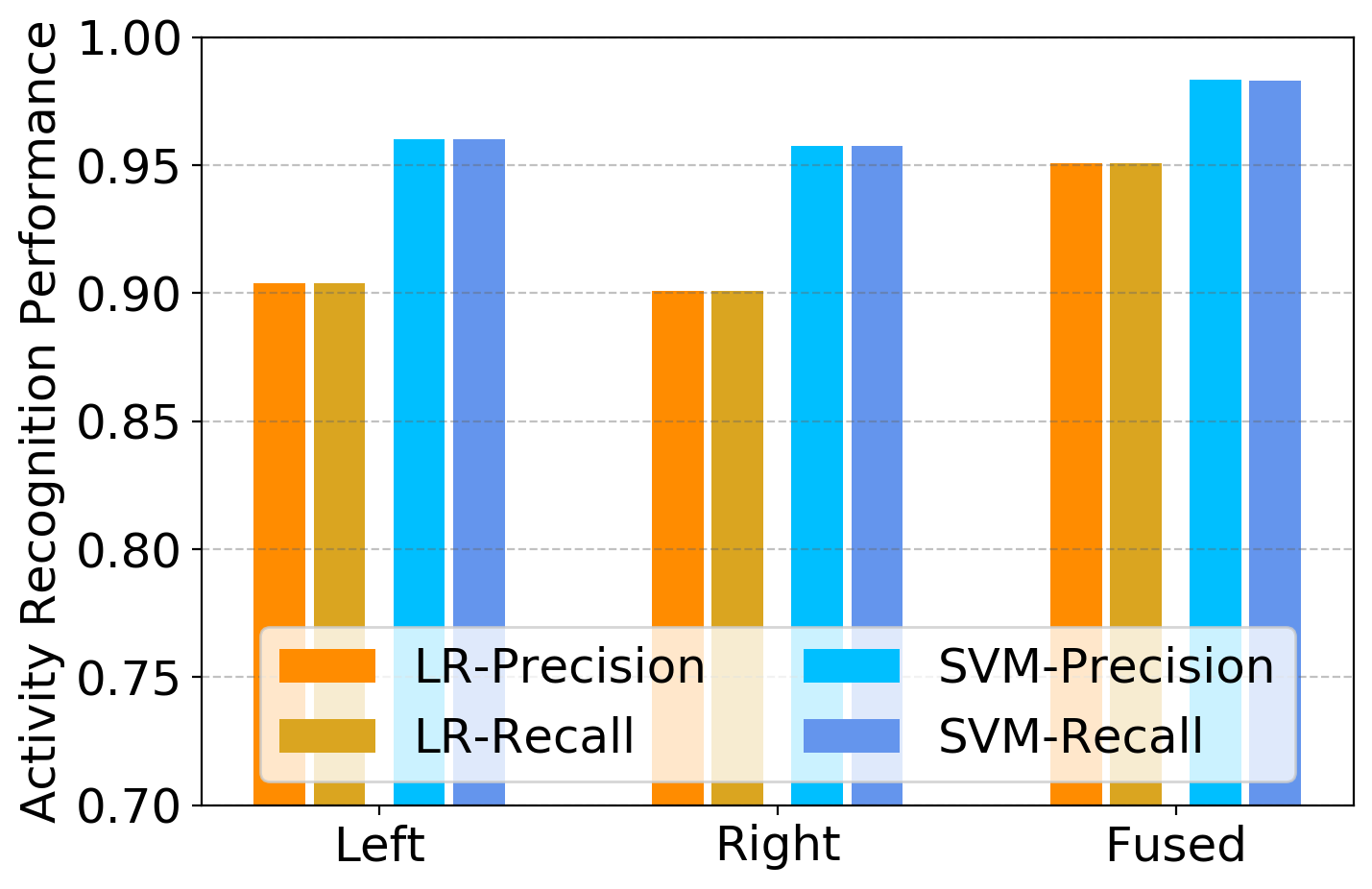}
		\label{fig:har_overall}}
	\subfigure[]{
		\centering
		\includegraphics[scale = 0.22]{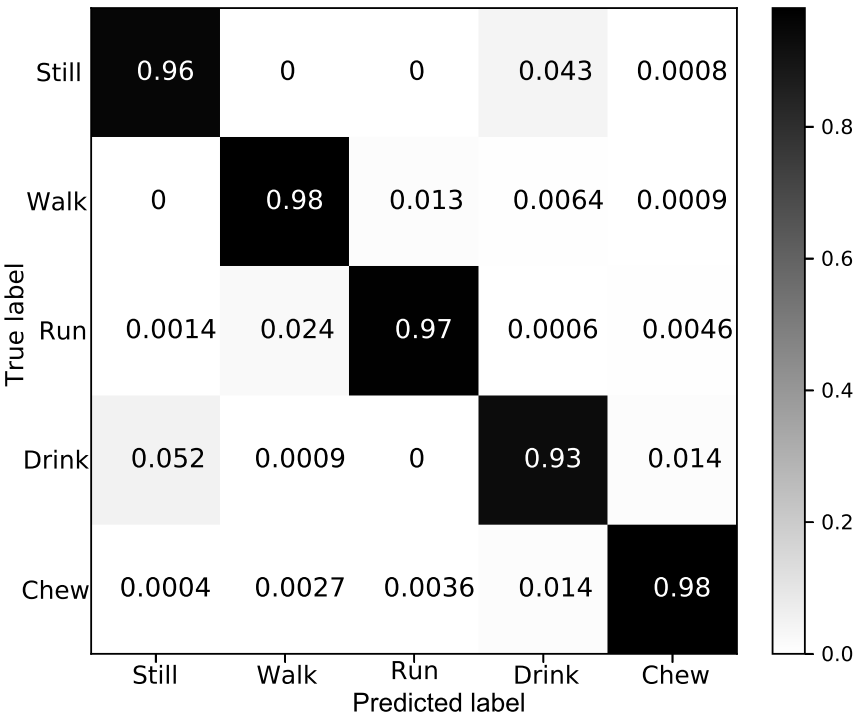}
		\label{fig:har_conf_mat}}
		\vspace{-0.2in}
	\caption{(a) Overall activity recognition performance, (b) Confusion matrix.}
	\vspace{-0.2in}
\end{figure}

{\bf Overall Performance:}
We train the individual model for each subject and compare the performance of three datasets: left earbud, right earbud, and fused (i.e., concatenate left and right earbud).
For each dataset from each subject, we split the data into the training set (80\%) and testing set (20\%), and train the models with 5-fold cross-validation. The results are averaged over all the subjects as shown in Figure~\ref{fig:har_overall}. We can observe that (1) SVM always achieves better performance than LR; (2) Left and right earbud achieve similar performance; (3) The fused dataset obtains the highest recognition precision and recall both of around 98.3\%. Such improvement might arise from the fact that the fused dataset gains from two sensing channels and is more resilient to signal distortions when one of the ear tips is loose. Therefore, we only present the results for SVM and fused dataset in the rest of the evaluation.

Figure~\ref{fig:har_conf_mat} illustrates the confusion matrix of the five activities using SVM. We can see that the trained model recognizes walking, running, and chewing quite well, while the main accuracy loss comes from the actions of being still and drinking being confused. This is originated from our data collection procedure: for drinking, the participants were obviously unable to swallow for prolonged periods continuously and the traces exhibit a resting period between swallowing episodes. If the rest period is longer than one second (window size), the recorded segment wound be similar to that of someone sitting still as no action is performed, resulting in confusion between drinking and being still. 

\begin{figure*}[ht]
\centering
	\subfigure[]{
		\centering
		\includegraphics[scale = 0.249]{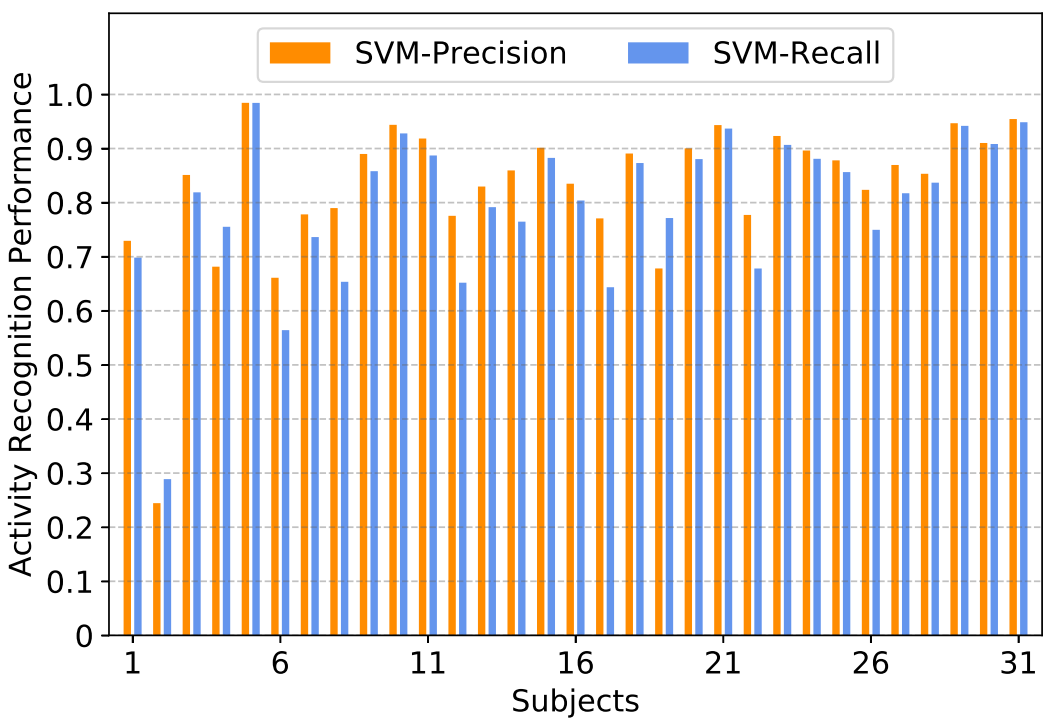}
		\label{fig:har_loo}}
	\subfigure[]{
		\centering
		\includegraphics[scale = 0.249]{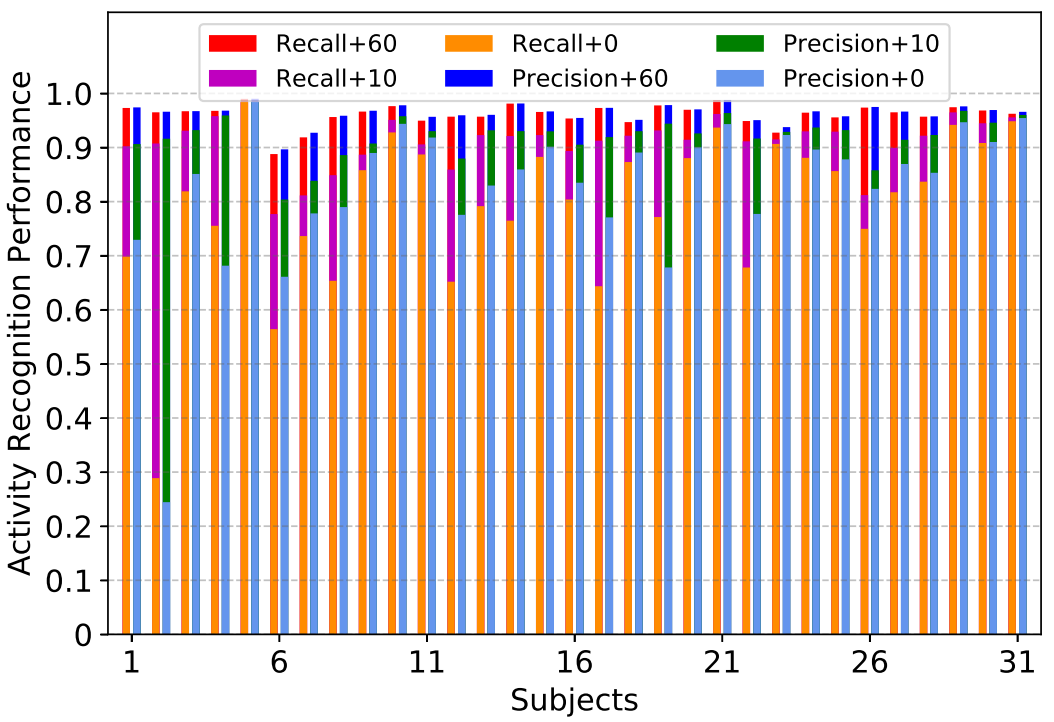}
		\label{fig:har_perfonalization_compare}}
	\subfigure[]{
		\centering
		\includegraphics[scale = 0.249]{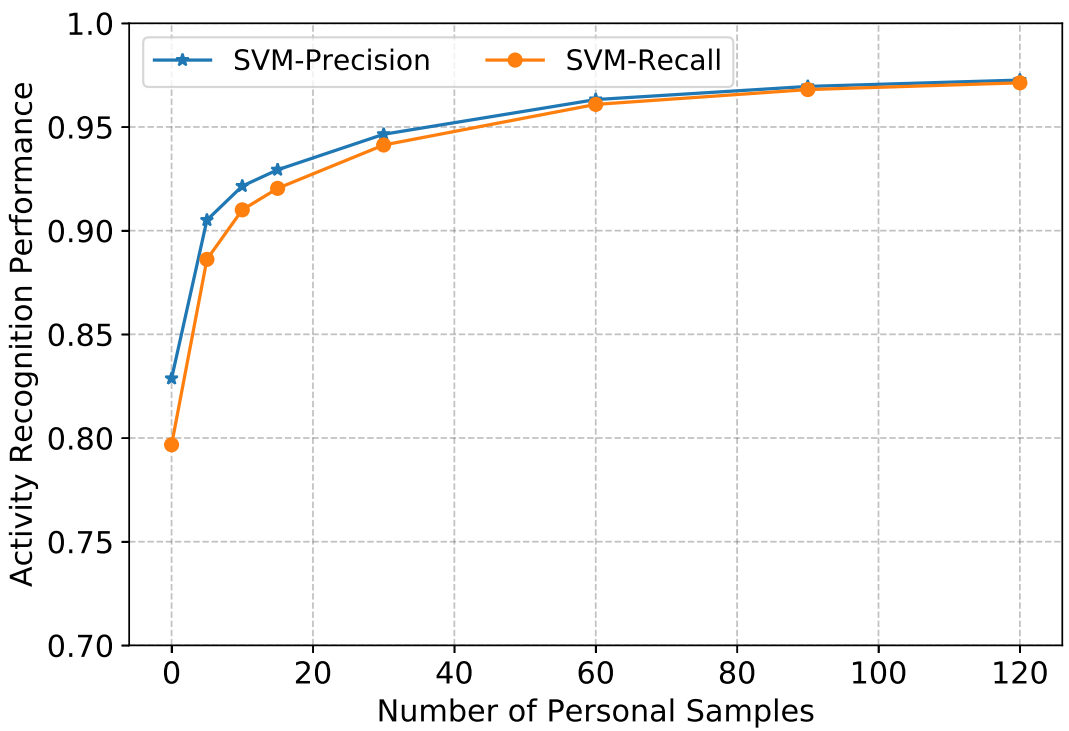}
		\label{fig:har_personalization}}
		\vspace{-0.15in}
	\caption{(a) Activity recognition performance for leave-one-out test, (b) individual performance with model personalization, and (c) average recognition performance with different amounts of personal data.}
	\vspace{-0.2in}
\end{figure*}

{\bf Leave-One-Out Test:} 
We perform the leave-one-out test on the fused dataset to justify how a pre-trained model can be generalized to a new user. In specific, we iteratively select one subject for testing and train the model using the data from the other subjects (30) with SVM. Figure~\ref{fig:har_loo} shows the recognition precision and recall for the leave-one-out test. It can be observed that the performance varies significantly among different subjects. The pre-trained model can be generalized well to some subjects, while it suffers from significant performance drop on other subjects. For example, Subject 5 achieves approximately 98.4\% recognition recall, while the value for Subject 2 is only 24.4\%. The reason might arise from the fact that people perform the five activities differently. For instance, the walking style (i.e, gait) of each person is unique. And people chew gum in distinct ways, such as slow/fast chewing and gentle/ravenous chewing. Thus, if the dataset used for pre-training does not cover the style of a new user, the model will perform poorly. \begin{figure*}[ht]
\centering
	\subfigure[]{
		\centering
		\includegraphics[scale = 0.254]{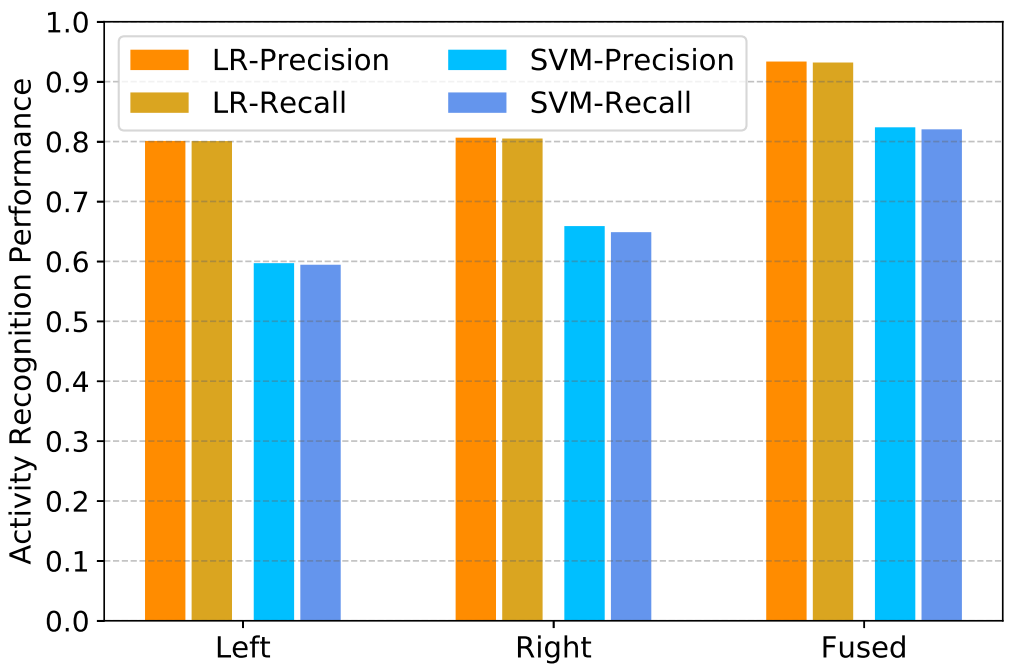}
		\label{fig:ges_overall}}
	\subfigure[]{
		\centering
		\includegraphics[scale = 0.254]{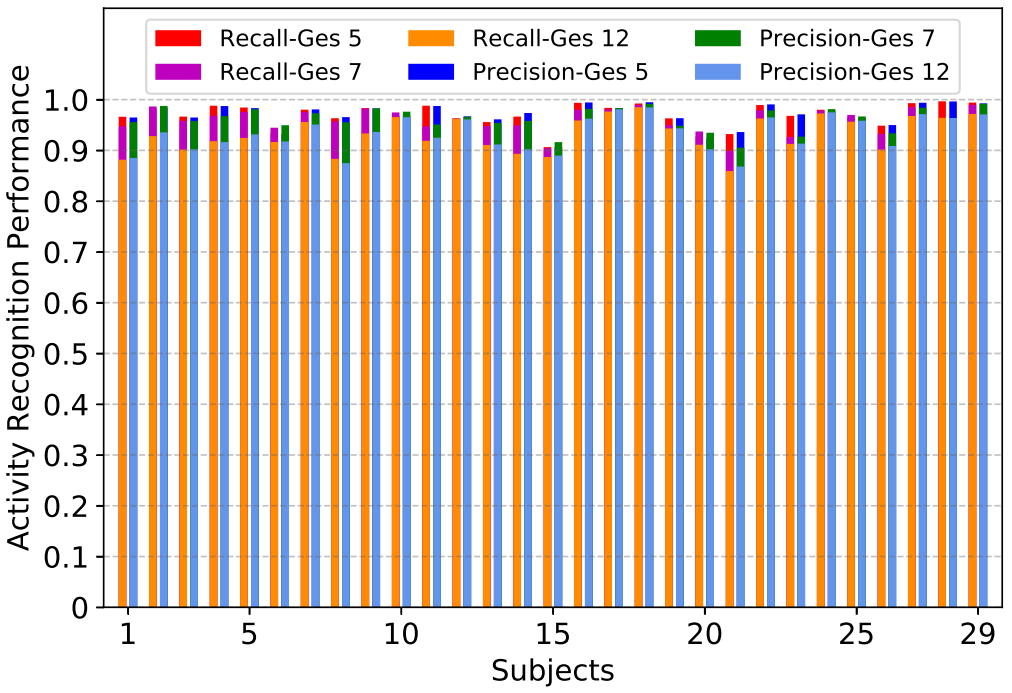}
		\label{fig:ges_gestureset_combine}}
	\subfigure[]{
		\centering
		\includegraphics[scale = 0.254]{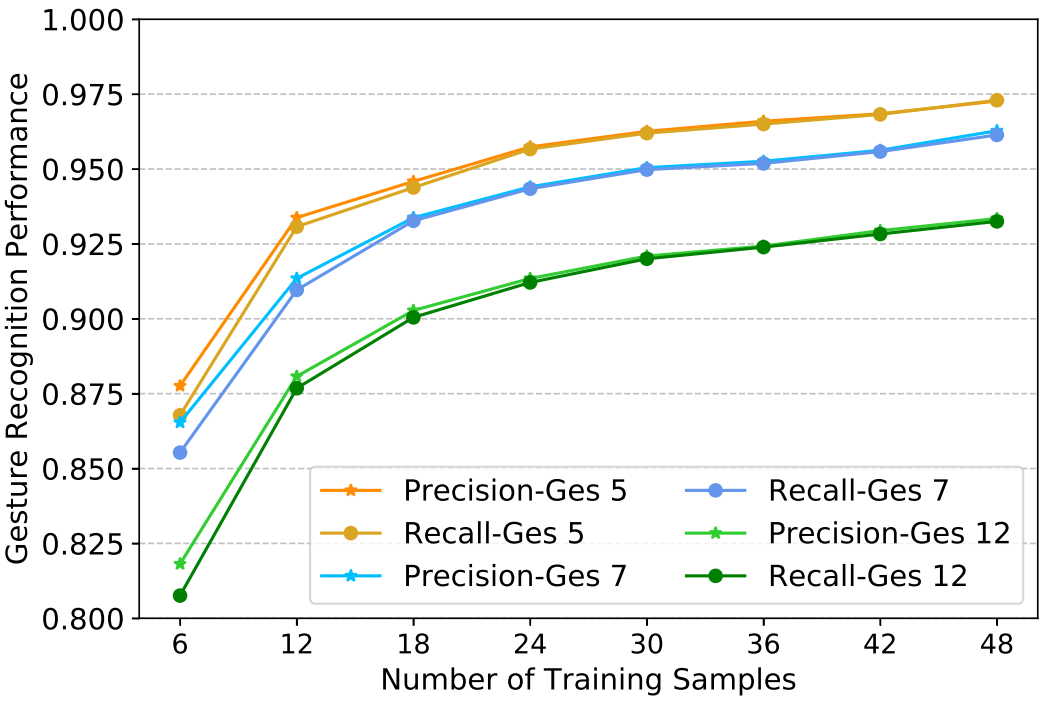}
		\label{fig:ges_training_size}}
		\vspace{-0.15in}
	\caption{(a) Overall gesture recognition performance (12 gestures), (b) individual gesture recognition performance with different gestures, (c) impact of training data size (average over 29 subjects).}
	\vspace{-0.1in}
\end{figure*}

{\bf Model Personalization:} 
A simple way to address the model generalization issue is by collecting activity data from many subjects (e.g., hundreds or thousands), so that the pre-trained model covers large variations of activity styles. However, this might lead to a tremendous burden on data collection. Here we explore an alternative technique that uses a user-specific model (model personalization) by re-training the general model with personal data, which only requires a limited amount of user data. For each iteration in the leave-one-out tests, we include different amounts of data from the testing subject for training and test on the rest. 

Figure~\ref{fig:har_perfonalization_compare} shows the recognition performance when 0, 10, and 60 samples from each subject are involved to re-train the model. We can observe that, with 10 samples, both the recognition precision and recall are significantly improved, especially for those perform poorly without personal data (e.g., Subject 2 improves from 24.4\% to 91.6\%). With 60 samples, the performance can be further enhanced.
Figure~\ref{fig:har_personalization} compares the average value with [5, 10, 15, 30, 60, 90, 120] personal samples added for model re-training. The results indicate that higher accuracy improvement can be achieved when more personal data is provided. The average precision reaches 92.1\% (enhances by 9.2\%) with 10 personal samples and 96.3\% (enhances by 13.4\%) when 60 personal samples being added. Targeting on 90\% precision, users only need to provide 10 samples for each activity, which can be collected within 25 seconds (5 seconds each) with window overlapping of 50\%. 
The requirement of user intervention (providing personal data) has been socially accepted in different applications, such as user authentication (profile registration using face or fingerprint) and IMU-based motion tracking (user-assisted sensor calibration with magnetometers~\cite{camps2009numerical}).

\begin{figure*}[ht]
\centering
	\subfigure[]{
		\centering
		\includegraphics[scale = 0.25]{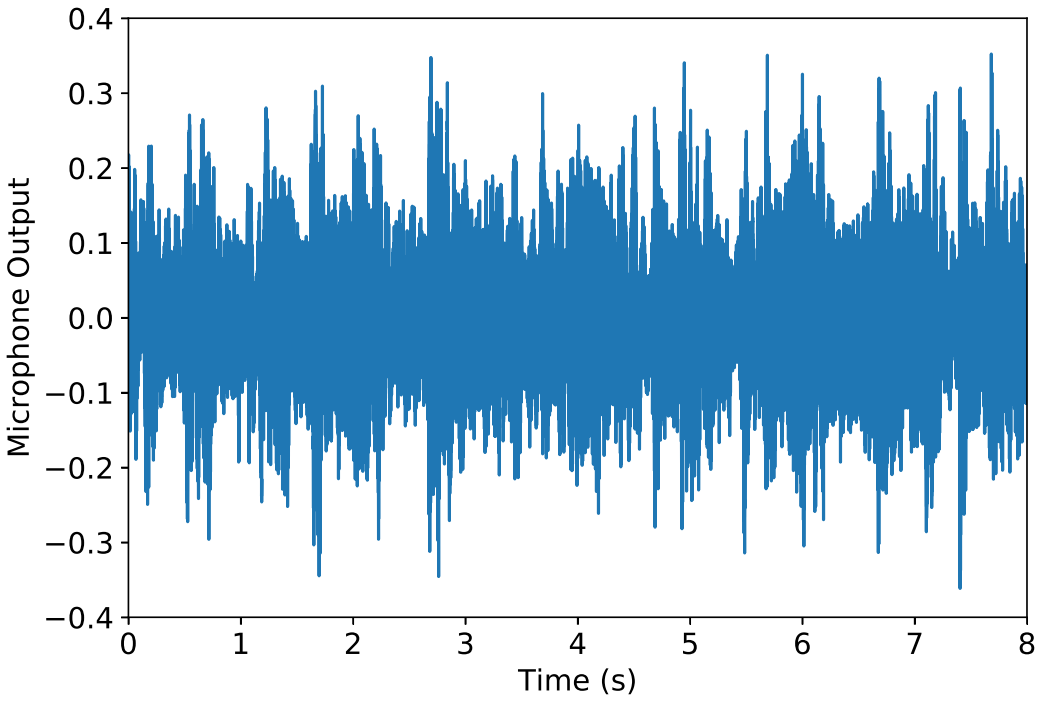}
		\label{fig:music_original}}
	\subfigure[]{
		\centering
		\includegraphics[scale = 0.25]{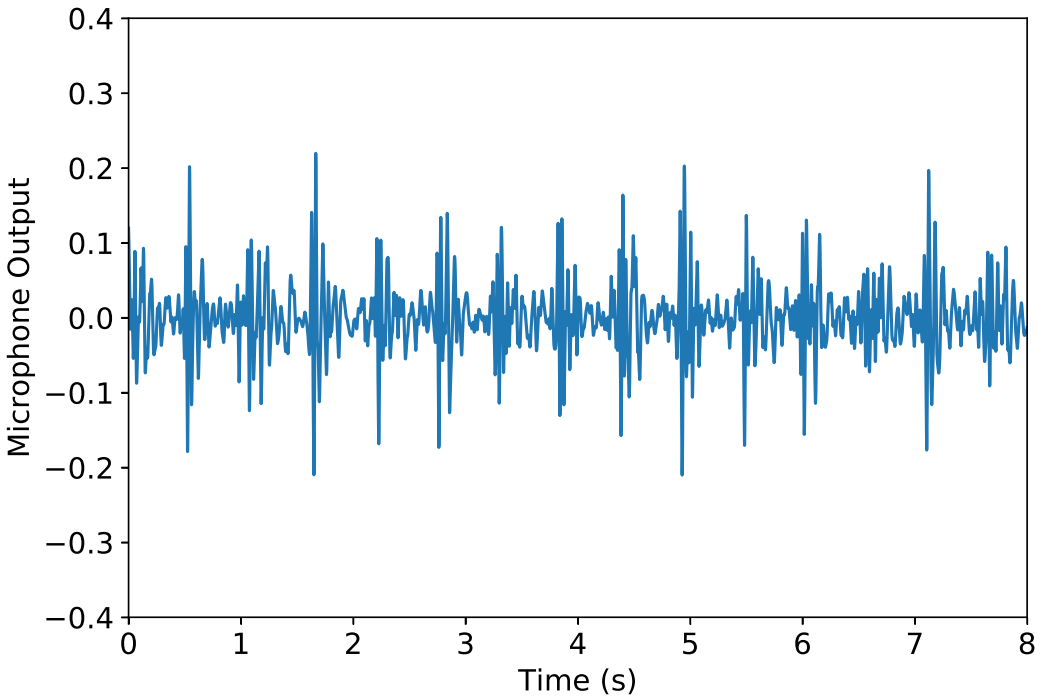}
		\label{fig:music_filtered}}
	\subfigure[]{
		\centering
		\includegraphics[scale = 0.25]{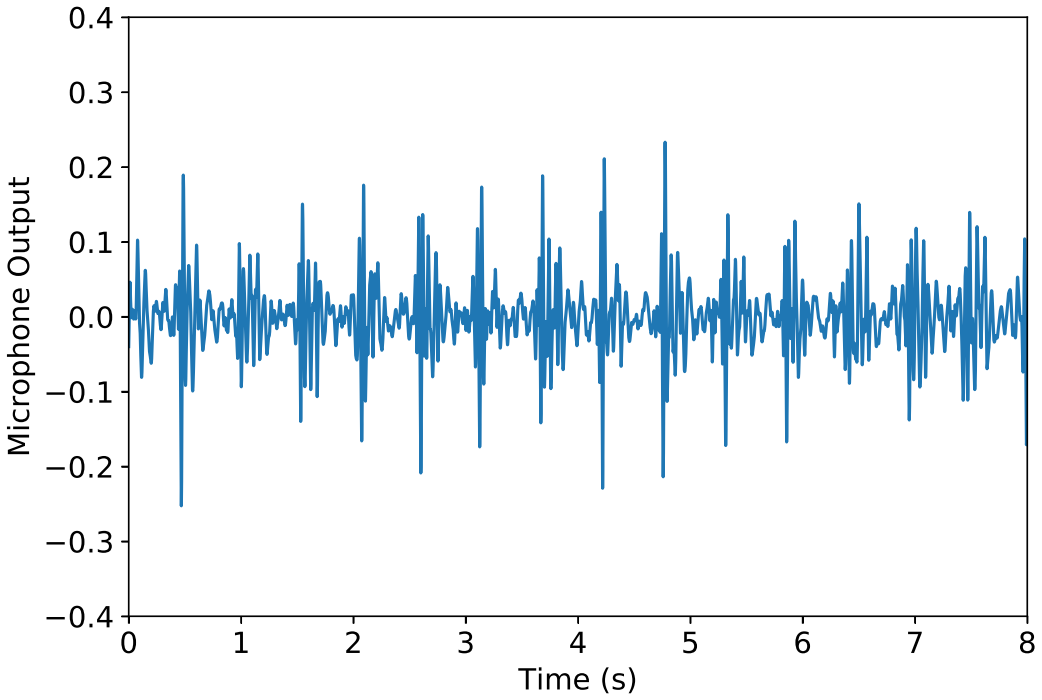}
		\label{fig:music_baseline}}
	\vspace{-0.2in}
	\caption{The (a) original signal and (b) low-pass filtered signal for participant walking during music playback, (c) low-pass filtered signal for the same participant walking without music playback.}
	\vspace{-0.1in}
\end{figure*}

\subsection{Hand-to-Face Gesture Recognition}
{\bf Overall Performance:}
Figure~\ref{fig:ges_overall} compares the gesture recognition performance (averaged over 29 subjects) among the two classifiers for the three datasets. We can see that LR consistently achieves better performance, which might be because the features from different gestures are likely to be separated linearly. As expected, the fused dataset (93.2\% recall) outperforms the two individual datasets (80.1\% and 80.5\% recall for left and right, respectively) on the 12 gestures, demonstrating the benefits of sensing with both earbuds. In addition, achieving 93\% recall over 12 gestures clearly demonstrates the quality of the sensing data. We then limit the classifier to LR and the dataset to fused version in the following evaluation.

\begin{figure*}[ht]
	\centering
	\includegraphics[scale = 0.53]{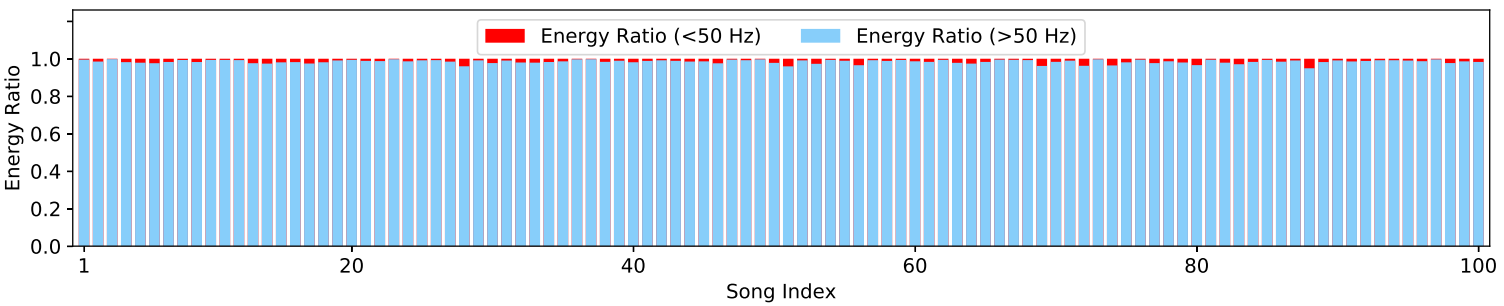}
	\vspace{-0.15in}
	\caption{Spectrum energy analysis of 100 songs.}
	\vspace{-0.15in}
	\label{fig:music_song_100}
\end{figure*}

{\bf Gesture Set Optimization:} 
Next, we perform gesture set optimization for the purpose of (1) further improving the recognition accuracy and (2) reducing the overhead for users to memorize gestures. In specific, based on the confusion matrix on the 12 gestures and spatial position of tapping spots, we narrow down the number of gestures to 7 and 5 by removing gestures with large confusion errors and with close proximity. The selected 7 gestures are \{\textit{Forehead Middle, Nose, Chin, Jaw Left, Jaw Right, Cheek Left, Cheek Right}\}. And the optimal 5 gestures are \{{\em Forehead Middle, Nose, Chin, Jaw Left, Jaw Right}\}. We then run the experiment with LR classifier on the two new gesture sets and Figure~\ref{fig:ges_gestureset_combine} compares the results for each individual. We can observe that although the recognition performance varies among subjects, most of them achieve >90\% precision and recall with 12 gestures. The performance improves further with gesture set optimization. Overall, the average recall for 12, 7, and 5 gestures is 93.2\%, 96.0\%, and 97.0\%, respectively.

{\bf Impact of Training Size:} 
The proposed hand-to-face gesture interaction is founded on the fact that vibrations from different tapping spots experience distinct paths to the ears: the model is actually trained to recognize these paths. Given that people have different head size, skull structure, and tissue composition, it is expected that the model trained on one subject could not be fitted to others. To confirm this, we run the leave-one-out test and obtain an average recognition recall of 20\%. As a consequence of this finding, it is clear that users should train a personal model with their own gesture data. To investigate the user burdens for training data acquisition, we re-train the model with different number of gesture samples. As shown in Figure~\ref{fig:ges_training_size}, the accuracy of the three gesture sets shows similar patterns with increasing training samples. With 12 samples from each gesture, the 5 gestures set obtains a recognition recall of 93.5\%. Based on the proposed data collection protocol (one gesture per second), only one minute is required. Also, providing user data to train the personal model is acceptable in other input/gesture interaction systems~\cite{chen2018vitype}, as long as the burden is small.

\subsection{Impact of Music Playback}
Given that the original functionality of earbuds is to deliver sounds (e.g., music and phone call) to the human ear, a common question is that will these sounds (usually much higher volume) pollute the audio sensing signals? To investigate this, we asked one participant to walk while the earbuds are playing a song with the built-in speakers at an appropriate volume. Figure~\ref{fig:music_original} illustrates the original signal collected from the left earbud and we can see it is dominated by the music. Figure~\ref{fig:music_filtered} shows the low-pass filtered (<50~Hz) version of the signal, where the steps can be clearly observed. Then, without music playing, the subject walked another trace in the same condition. The signal after low-pass filtering is plotted in Figure~\ref{fig:music_baseline}. Visually, we can see that the two filtered versions have high similarity and the step counts can be easily derived. We also quantify the similarity of signals from frequency domain using the structural similarity (SSIM, a well-known metric to compare similarity between two images~\cite{sampat2009complex}). Specifically, we first obtain the spectrogram of each signal using short-time Fourier transform (STFT), and then calculate the SSIM index~\footnote{SSIM index ranges from 0 to 1. The higher the value, the greater the similarity between the images.} between two spectrograms (images). Our results show that the SSIM index for \{Figure~\ref{fig:music_original}, Figure~\ref{fig:music_filtered}\}, \{Figure~\ref{fig:music_original}, Figure~\ref{fig:music_baseline}\}, and \{Figure~\ref{fig:music_filtered}, Figure~\ref{fig:music_baseline}\} is 0.35, 0.33, and 0.95, respectively, suggesting that music playback has extremely limited impacts.

To confirm that the performance of \SysName is robust against different types of music, we analyze the spectrum of the All-time Top 100 Songs launched by Billboard~\cite{top100songs}. For each song, we perform the fast Fourier transform (FFT) analysis and obtain the signal energy at each frequency band. Then, we sum up the energy with frequencies below and above 50~Hz and calculate the energy ratios of the two frequency ranges, respectively. As shown in Figure~\ref{fig:music_song_100}, the energy of all the songs is dominated by frequencies higher than 50~Hz. The average energy ratio of <50~Hz signal is only 1.5\%, indicating that the impact of music playback is negligible. We also superimpose the music signals on the activity signals (i.e., sample-wise summation), which produces signals containing both low-frequency activity sound and high-frequency music. Then, a low-pass filter (<50Hz) is applied to the signals and we calculate the Pearson correlation between the low-pass filtered signal and the original activity signal. The correlation is about 0.982, further proving the robustness of OESense to music.

To further demonstrate this in practice, we collected the activity data from one subject when the earbuds were playing music with the built-in speakers. 
We then applied the proposed signal processing and classification pipeline to the collected data. The results show that OESense achieves 97.5\% recognition recall. 
In terms of phone calls, the frequency range of human voice over telephony transmission is within 300-3400~Hz\cite{esteban19789}, so that it can be completely removed after the low-pass filtering. For low frequency noise in the environment (e.g., fan motion), OESense leverages the sealing of the ear canal, which serves as an additional layer of filter to further suppress the noise, making the internal microphone less vulnerable to external sounds (as presented in~\Cref{sec:occ_effect}). 

\begin{table*}
\centering
\begin{minipage}{0.28\linewidth}
\centering
\caption{In-the-wild step counting accuracy and recognition recall.}
\vspace{-0.1in}
\label{tab:field}
\begin{tabular}{lrrr}
\toprule
\textbf{Subject} & \textbf{SC} & \textbf{AR} & \textbf{GR}  \\ \cmidrule{1-4}
Subject 1 & 95.14\% & 90.60\% & 95.12\% \\ \cmidrule{1-4}
Subject 2 & 97.33\% & 92.08\% & 93.44\% \\ \cmidrule{1-4}
Subject 3 & 96.97\% & 88.95\% & 92.17\% \\ 
\bottomrule 
\end{tabular}
\end{minipage}
\hspace{0.1in}
\begin{minipage}{0.37\linewidth}
\centering
\caption{Power consumption of OESense for gesture recognition.}
\vspace{-0.1in}
\label{tab:power}
\begin{tabular}{lr}
\toprule
\textbf{Operation} & \textbf{Power (mW)} \\ \cmidrule{1-2}
RasPi(Baseline) & 2,340 \\ \cmidrule{1-2}
RasPi+MicRecd & 2,459 \\ \cmidrule{1-2}
RasPi+MicRecd+GesRecg & 3,086(LR), 3,106(SVM) \\ 
\bottomrule 
\end{tabular}
\end{minipage}
\hspace{0.1in}
\begin{minipage}{0.28\linewidth}
\centering
\caption{Latency of OESense for gesture recognition.}
\vspace{-0.1in}
\label{tab:latency}
\begin{tabular}{lr}
\toprule
\textbf{Operation} & \textbf{Latency (ms)} \\ \cmidrule{1-2}
Low-pass Filter & 1.54 \\ \cmidrule{1-2}
Feature Extract & 38.97 \\ \cmidrule{1-2}
Inference & 0.34(LR), 0.95(SVM) \\
\bottomrule 
\end{tabular}
\end{minipage}
\vspace{-0.1in}
\end{table*}

\subsection{In-the-Wild Study}
\label{sec:field}

In this section, we sought to assess the in-the-wild performance of OESense under more realistic settings. To do so, we recruited three additional subjects (one female, two males). One subject was at home, while the other two subjects at their workplace. The subjects wore the device and performed the three applications separately for a total duration of about one hour. To assess the performance of OESense in counting steps, the subjects were free to walk or sit down. On the other hand, for activity and gesture recognition, we had to first collect a one-minute sample for each activity/gesture to train the model. The subjects then performed the activities/gestures freely, in random order. In compliance with our institution’s approved ethics application, a camera was used to record video as ground truth.

Then, we analyze the data with the proposed signal processing and recognition pipelines (LR as the classifier). The results are presented in~\Cref{tab:field}. For step counting, the accuracy drops by 2\%-4\% compared to the controlled setting. We observed that the accuracy variance for the three subjects stems from the way they walk: our approach tends to perform well if a user walks continuously for a certain period (e.g., longer than one minute), whereas it struggles to count choppy strides or light steps. For activity and gesture recognition, we observe a slightly higher accuracy drop. This might be due to the fact that there is variance among how people perform the actions and the trained model cannot generalize well due to the limited training samples. However, OESense still achieves more than 90\% accuracy for the three applications, demonstrating its effectiveness in practical scenarios.

\subsection{Power and Latency Measurement}
\label{sec:power}
To the best of our knowledge, there is no open platform to instrument a stand-alone earable sensing system. As a matter of fact, most of today's earables are directly offloading their data either to a phone or via the network.
Instead, here, we wanted to explore the system-level performance of OESense as a stand-alone system, as we envisage some of the future earables will become stand-alone. We opted for a Raspberry Pi 4B as a reference platform.
Taking hand-to-face gesture recognition as an example, we evaluate the power consumption and latency of our system. 
We train the machine learning models (LR and SVM) on a laptop and implement the gesture recognition pipeline (including low-pass filtering, feature extraction, and inference) on the Raspberry Pi. As shown in~\Cref{tab:power}, compared to baseline (idle) power consumption (2,340~mW), powering and recording microphone data (MicRecd) consumes an additional 119~mW. 
This value is dramatically greater than the power draw of the microphone (360~$\mu W$) reported in the datasheet~\cite{spu1410}, which indicates that most of the power is consumed by the Raspberry Pi for data sampling. 
When gesture recognition is performed concurrently, the Raspberry Pi consumes 746~mW and 766~mW overall for LR and SVM, respectively. 
Although at a first sight these figures may seem substantial, it is worth noticing that (1) unlike low-power microprocessors (especially those designed for audio processing, like the Apple H1 chip embedded on the AirPods Pro), the Raspberry Pi is known to be power-hungry without energy efficiency optimization; (2) recognition (feature extraction and inference) is run only whenever a gesture is detected and such operation time is typically very short, so the actual energy consumption would be much lower. \Cref{tab:latency} shows the operation time of procedures in the recognition pipeline. The majority of the time is due to feature extraction (38.97~ms), while inference time is almost negligible (0.34~ms for LR and 0.95~ms for SVM), granting OESense quasi-real-time performance.

To further ground our study with practical considerations, we estimated the overhead of running gesture recognition over any other earbud functionality, for a possible worst case scenario -- a user is performing a gesture (0.4~s long) every 2~s , a very aggressive assumption. 
The average energy consumption per second would be $E = 119mW \times 1s + 627mW \times 40.85ms \times 0.5 = 131.8mJ$.
Considering a wireless earbud with a battery like that of an AirPod Pro (81~mAh), OESense could operate for a time $T = \frac{81mAh \times 5V}{131.8mJ} = 3.07h$. We also measured the power consumption of the proposed system in the wild. 
To do so, we used a power bank of 20,000~mAh; our measurements show that OESense can continuously operate for more than 15 hours.
Although these are ballpark figures, on a non-optimized off-the-shelf device, they give an indication of the actual feasibility of OESense in practice.

\section{Discussion and Limitations}
\label{sec:dicussion}

During OESense prototyping, we changed the position of the speaker to embed the microphone. Such design optimizes the SNR of the microphone, whereas modifies the internal structure of the earbud. Thus, we assess whether the audio quality of music playback is affected through a user study. Each participant listens to a music segment with unmodified and modified earbuds respectively and rates their perception towards the audio quality. 29/31 subjects reported that no difference is perceived between the two earbuds and 2 subjects even reported that the modified earbud has slightly better audio quality. So the add-on sensing capability would not impact the audio delivery quality. However, there are also several limitations.

First, we only demonstrated the concept of \SysName with in-ear headphones only due to the requirement of occluding the ear canal opening. Such physical occlusion might lead to impaired awareness of the surrounding environment (e.g., traffic sounds) and incur safety issues. A possible solution is to imitate the transparency mode on AirPods Pro devised by Apple~\cite{airpodspro}. In specific, the external microphone can measure the outside sounds and replay the meaningful parts (like sirens, between 725-1600~Hz~\cite{angione2016study}) through the onboard speakers. Besides, wearing in-ear earbuds for a long time might cause discomfort to the user. Given the occlusion effect is present as long as there is a sealed gap between the occlusion device and the person's eardrum, we leave as a future work an exploration of the performance of OESense using over-the-ear headphones, which would provide another alternative form factor of OESense to suit user preference.

Second, we implemented the concept of OESense on a Raspberry Pi based data collection and processing system, which is cumbersome and impractical for mobile scenarios. Besides, the Raspberry Pi consumes substantial power (\Cref{sec:power}) as it is not optimized for low-power applications. Thus further efforts to implement OESense in an energy-efficient manner are required. With advanced audio chips (like Apple H1 chip) and dedicated PCB design, the power consumption could be significantly reduced.
In addition, current \SysName prototype is built upon a pair of wired earbuds, where the connection wires swing during movements and produce additional noise. This impact is expected to disappear with wireless earbuds. However, data synchronization between two ears should be considered when wireless earbuds are adopted.

Third, while we have demonstrated that OESense can detect the three sensing applications separately, whether it is able to run these concurrently remains unclear.
For example, from \Cref{fig:step_counting_algorithm} and \Cref{fig:ges_waveforms}, we can observe that the signal waveforms of face taps and walking look quite similar. 
If a face tap happens exactly at the same time as a foot strike, OESense may not be able to distinguish whether it is a step or a gesture.
However, if they are not perfectly overlapping, there are still ways to differentiate them.
Concretely, a tap signal usually has fewer spikes than a step signal, therefore, the zero-crossing rate can be a potential metric. 
With the collected data, we calculated the zero-crossing rate for different gestures and steps. 
The results show this approach can correctly recognize 97.1\% samples. We also tried a model-based method: by training a k-nearest neighbors (KNN) model to recognize tap gestures and steps, we achieve 96.3\% accuracy.

A further limitation of OESense is that it is unable to detect a variety of very light activities, e.g., humans horizontally shaking their heads slowly. 
Instead, it can only sense motions that generate vibrations propagating through the 
human body.
Additionally, the occlusion effect is established in the small cavity between the ear-tip and the eardrum. It is unclear how obstacles inside the ear canal would affect the occlusion effect. Thus, the impact of otitis media (i.e., fluid effusion in the middle ear) on OESense performance might need further investigation.

\balance
\section{Related Work}

{\bf Sensing with Earables:}
Health monitoring with earables has recently attracted intensive attention. By incorporating the photoplethysmography (PPG) sensor into an earbud, Poh et al.~\cite{poh2009heartphones} proposed to measure heart rate as ear blood vessels reflect different amounts of light during a heartbeat. With the combination of PPG and an inflation-controllable balloon, Bui et al.~\cite{bui2019ebp} built a system to unobtrusively estimate blood pressure from the artery in the ear canal.  
Martin et al.~\cite{martin2017ear} designed an earbud to measure the pressure in the occluded ear canal and estimate heart rate based on the pressure variations.
For motion detection, researchers mainly explored facial expression recognition and activity recognition with various modalities. Ando et al.~\cite{ando2017canalsense} observed that the human ear canal is deformed distinctively among facial expressions and proposed to recognize them by capturing the resulted pressure change with a barometer. Differently, Matthies et al.~\cite{matthies2017earfieldsensing} utilized the electrodes on earbuds to capture the electricity variations caused by muscle movements during expressions. Using the eSense platform's IMU, Prakash et al.~\cite{prakash2019stear} proposed to count human steps and Ferlini et al.~\cite{ferlini2019head} investigated head motion tracking.

{\bf Acoustic-based Activity Recognition:}
Due to the proliferation of microphones and speakers on Internet of Things and wearable devices, acoustic sensing has been widely adopted in a broad range of applications. Various properties of audio signals, such as time-frequency features~\cite{wang2014ubiquitous,ren2015fine,chauhan2017breathprint}, time difference of arrival (TDoA)~\cite{liu2015snooping}, Doppler shift~\cite{gupta2012soundwave}, phase~\cite{wang2016device}, and channel impulse response~\cite{yun2017strata}, have been exploited for motion sensing. Based on whether a speaker is involved to generate a reference audio beacon, these applications can be categorized into passive sensing and active sensing. In the interest of space, we focus on passive acoustic sensing as our work belongs to this category. 

Passive sensing measures the sound generated by target activities/motions with microphones. Wang et al.~\cite{wang2014ubiquitous} presented a keystroke recognition system named UbiK using the smartphone microphone. Ubik calculates the amplitude spectrum density (ASD) of the clicking sound and identifies different keystrokes using a fingerprinting-based method. Employing dual-microphones design on current smartphones, Liu et al.~\cite{liu2015snooping} proposed to discriminate keystrokes based on the TDoA of the keystroke sound at the two phone microphones. By recording the sound during sleeping, Ren et al.~\cite{ren2015fine} detected human breathing rate with a correlation-based method and recognized different events (like cough and snore) with the Mel-frequency cepstral coefficients to estimate the quality of sleep. Chauhan et al.~\cite{chauhan2017breathprint} proposed to authenticate users using the breathing sound.

{\bf Hand-to-Face Interaction:}
As the human face offers a large area for interactions with head-worn devices, hand-to-face gestures have been explored using various modalities. Serrano et al.~\cite{serrano2014exploring} proposed to detect hand-to-face gestures using a camera array. Kikuchi et al.~\cite{kikuchi2017eartouch} built an earbud embedded with four photo-reflective sensors to measure the deformation of ear rim. Also, Lee et al.~\cite{lee2017itchy} used an Electrooculography (EOG) sensor equipped on the nose pad of commercial eyeglasses to detect nose touching gestures.

Close to our work, Xu et al.~\cite{xu2020earbuddy} presented a hand-to-face interaction system called EarBuddy using acoustic signals. However, EarBuddy recognizes face-touching gestures using audible air-conducted gesture sound recorded by an external microphone on the earbud body, which suffers from low signal-to-noise ratio (SNR) due to the dramatic attenuation of sound in the air (\Cref{sec:explore_emic}). As a result, EarBuddy can only recognize gestures close to the ear with good performance. Further, it suffers from acoustic noise and can work in the quiet environment only. In contrast, our method takes advantage of the occlusion effect and leverages an in-ear microphone to record bone-conducted sounds, which not only improves the SNR but is also immune to external noise. In addition, compared to tapping on earbuds directly, the human face can provide a larger input interface.

\section{Conclusion}

We presented OESense, a novel human sensing system based on audio signals recorded inside the ear. Leveraging the occlusion effect, \SysName shows great sensing potential for both intense and light human activities. We demonstrated three sensing applications (i.e., step counting, activity recognition, and hand-to-face gesture interaction) with the developed \SysName prototype. All achieved good performance (average recall of 99.3\%, 98.3\%, and 97.0\%, respectively). Further, our system analysis suggests that OESense can achieve quasi-real-time performance with acceptable power consumption. Given that the ear contains abundant information of human signs and motions, \SysName is potential to be extended to more personal-scale sensing applications like heartbeat detection, jaw movement detection, facial expression recognition, and so on.
\vspace{-0.1in}
\section{Acknowledgments}

This work is supported by ERC through Project 833296 (EAR) and by Nokia Bell Labs through a donation. We thank D. Spathis for the insightful discussions, the anonymous shepherd and reviewers for the valuable comments, and the volunteers for the data collection.

\bibliographystyle{unsrt} 
\bibliography{main}
\balance
\end{document}